\pdfoutput=1 
\documentclass[floatfix, aps,prmaterials,twocolumn, groupedaddress]{revtex4-2}
\usepackage{dcolumn}
\usepackage{bm}
\usepackage{amssymb}
\usepackage{amsthm}
\usepackage[version=4]{mhchem}
\usepackage{graphicx}[dvipdfmx]
\usepackage{natbib}
\usepackage{subfig}
\usepackage{booktabs}
\usepackage{multirow}
\usepackage[T1]{fontenc}
\newcommand{\degC}{$^\circ$C}
\newcommand{\etal}{\textit{et al}.}

\newcommand{\Biso}{B_{\rm iso}}
\newcommand{\TC}{T_{\rm C}}
\newcommand{\TN}{T_{\rm N}}
\newcommand{\TCC}{T^*_{\rm C}}
\newcommand{\TCW}{\theta_{\rm CW}}
\newcommand{\peff}{p_{\rm eff}}
\newcommand{\uB}{\mu_{\rm B}}
\newcommand{\kB}{k_{\rm B}}
\newcommand{\Msat}{M_{\rm sat}}
\newcommand{\Haniso}{H_{\rm aniso}}
\newcommand{\Hmone}{H_{{\rm m}_1}}
\newcommand{\Hmtwo}{H_{{\rm m}_2}}
\newcommand{\Hm}{H_{\rm m}}
\newcommand{\HE}{H_{\rm E}}
\newcommand{\HA}{H_{\rm A}}

\newcommand{\MAB}{\ce{$M$_2AlB_2}}
\newcommand{\xMAB}{\ce{(Fe_{1-$x$}Mn_{$x$})_2AlB_2}}
\newcommand{\FeMAB}{\ce{Fe_2AlB_2}}
\newcommand{\MnMAB}{\ce{Mn_2AlB_2}}
\newcommand{\CrMAB}{\ce{Cr_2AlB_2}}
\newcommand{\xMMAB}{\ce{(Fe, $M$)_2AlB_2}}
\newcommand{\xAMAB}{\ce{Fe_2(Al, $A$)B_2}}

\begin{document}


\title{Competing magnetic correlations and uniaxial anisotropy in {\xMAB} single crystals}


\author{Taiki Shiotani}
\email[]{Corresponding author: shiotani.taiki.48e@st.kyoto-u.ac.jp}
\affiliation{Department of Materials Science and Engineering, Kyoto University, Kyoto 606-8501, Japan}

\author{Takeshi Waki}
\affiliation{Department of Materials Science and Engineering, Kyoto University, Kyoto 606-8501, Japan}

\author{Yoshikazu Tabata}
\affiliation{Department of Materials Science and Engineering, Kyoto University, Kyoto 606-8501, Japan}

\author{Hiroyuki Nakamura}
\affiliation{Department of Materials Science and Engineering, Kyoto University, Kyoto 606-8501, Japan}

\date{\today}

\begin{abstract}
We have succeeded for the first time in synthesizing single crystals of nanolaminated borides {\xMAB} in the entire Fe--Mn composition range using the Al self-flux method, and have established $T$--$x$, $H$--$T$ and three-dimensional $H$--$T$--$x$ magnetic phase diagrams from the results of magnetization measurements.
The ferromagnetic correlation of {\FeMAB} is weakened with the Mn substitution, whereas the antiferromagnetic correlation of {\MnMAB} is enhanced with the Fe up to $x=0.65$. The spin direction in the magnetic ordered states changes from the $a$ to the $b$ axis with increasing Mn concentration and temperature.
At $x$ = 0.31--0.46, there are three magnetic phases; ferromagnetic, antiferromganetic, and intermediate phases in between.
At $x$ = 0.65 and 0.74, a spin-flop-like metamagnetic transition was observed at a finite field parallel to the spin direction. 
These observations indicate that in {\xMAB} the ferromagnetic and antiferromagnetic correlations coexist and the uniaxial magnetic anisotropy competes between the $a$ and $b$ axes.
\end{abstract}


\maketitle

\section{Introduction}
\begin{figure}[ht]
    \centering
    \includegraphics[keepaspectratio, width=1\columnwidth]{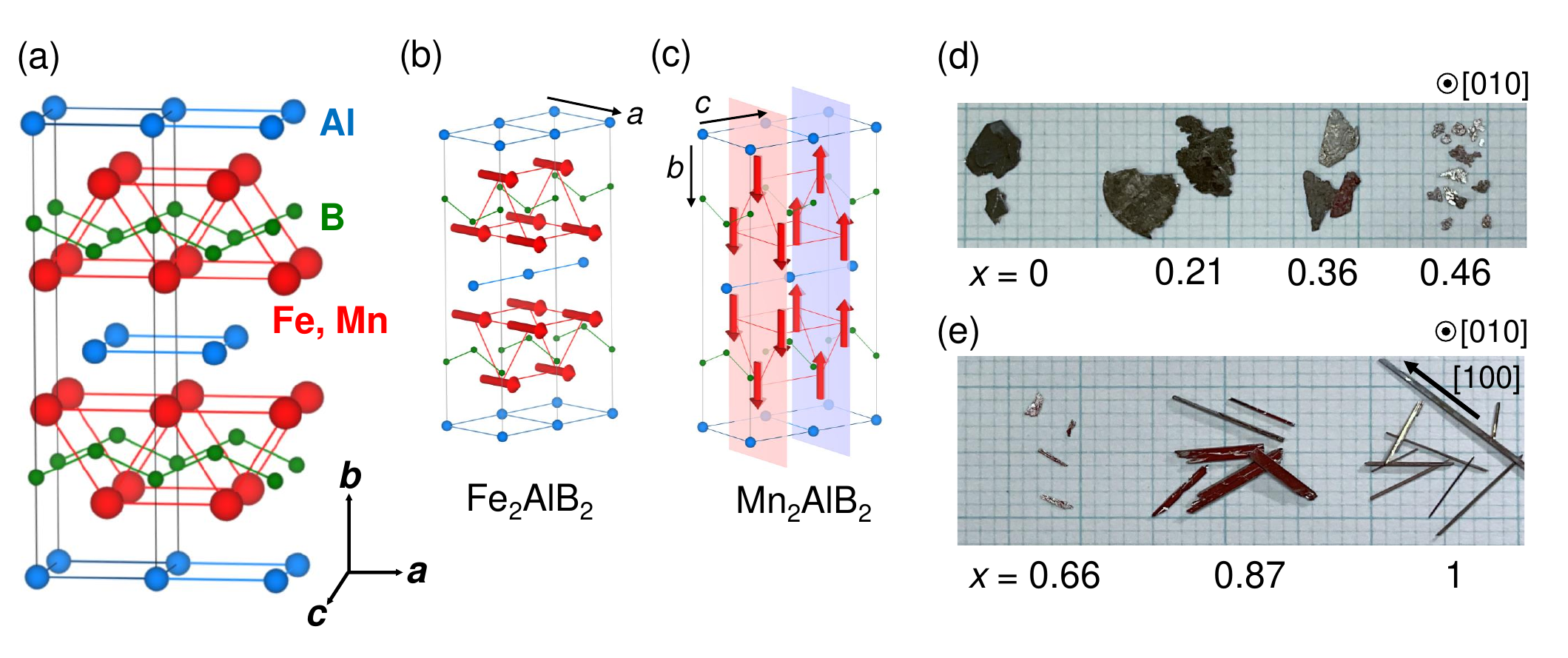}
    \caption{(a) Crystal structure of {\xMAB}. (b), (c) Magnetic structures of {\FeMAB} and {\MnMAB}, respectively, proposed in Refs. \cite{FeSCMAB, MnSCMAB, MABMnmagpowd} (d), (e) Image of grown single crystals of {\xMAB} with actual $x=0$--0.46 out of Al-self flux and those with $x=0.66$--1, respectively, on a mm scale. $x$ is actual Mn concentration measured by WDX.}
    \label{str}
\end{figure}

Transition-metal conductors with nanolaminated structure are of great interest as quasi-two-dimensional itinerant electron magnets in the viewpoints of both application and fundamental sciences. Since the discovery of the combination of metallic and ceramic properties, such as electrical and thermal conductivity, thermal and oxidative stability, and high hardness, in nanolaminated carbides and nitrides, called MAX phases \cite{MAX}, the design of magnetic properties has been intensively carried out to expand possible applications \cite{liuA, liuB, CrMnMAXa, CrMnMAXb}. However, near-room-temperature magnetism has not been realized in the bulk materials.
Recently, the nanolaminated transition-metal borides {\MAB} ($M=\ $Cr, Mn and Fe), which are members of a family called MAB phases, have attracted a lot of attention as the new candidates \cite{MABrev, newMABrev}. They form the orthorhombic structure in the space group $Cmmm$ (No. 65) with an alternating stacking of Al layer and $(M{\rm B})_2$ slab along the $b$ axis (Fig.\ \ref{str}(a)) \cite{Jeitschko, Kuzma}. The $(M{\rm B})_2$ slab has strong covalent boron-zigzag chains and metallic $M$--$M$ bonding, which leads to the coexistence of metallic and ceramic properties \cite{MABCrbonding}.
In addition to the characters of environment-resistant conductors, they exhibit itinerant magnetism of 3$d$ electrons in the $(M{\rm B})_2$ slab; {\CrMAB} is a Pauli paramagnet \cite{MABCrmag}. {\MnMAB} shows antiferromagetic ordering below the N\'{e}el temperature of $\TN = 313$\ K with spins aligned along the $b$ axis and the propagation vector of ${\bm q}=(0,0,1/2)$ (Fig.\ \ref{str}(b)) \cite{MnSCMAB, MABMnmagpowd}. {\FeMAB} is a ferromagnet with the Curie temperature of $\TC = 273\ $K, an ordered moment of $1.2\ \uB$/Fe, and the easy magnetization axis along the $a$ axis (Fig.\ \ref{str}(c)) \cite{FeSCMAB}. 
Density functional theory (DFT) calculations and angle-resolved photoemission spectroscopy revealed high density of states due to nearly flat bands of 3$d$ orbitals near the Fermi energy, which are associated with the magnetic ordering \cite{MABband, FeMABband, MnDOS}.\\
\indent Since the discovery of its excellent magnetocaloric properties near room temperature \cite{FeMABJACS}, {\FeMAB} has been extensively studied as a candidate for rare-earth-free and non-toxic magnetic refrigeration materials (for example, Refs. \cite{FeSCMAB, FeMABstrict, FeMABband, nonsto, FeMABmicrowave, FeMABstrchange, FeMABanisomagcalo, FeMABimproved, FeMABfirstorder,maganisoFe}). 
Detailed study using single crystal {\FeMAB} revealed substantial magnetocrystalline anisotropy with the anisotropy fields of 10 kOe along the $b$ axis and 50 kOe along the $c$ axis with respect to easy $a$ axis \cite{FeSCMAB}. Subsequently, the magnetocaloric properties were reported to depend on the direction of applied fields, and then the significant contribution of the magnetocrystalline anisotropy was confirmed, providing a new guideline for maximizing the magnetocaloric potential of {\FeMAB} \cite{FeMABanisomagcalo,maganisoFe}.
Investigations using the solid solutions {\xMMAB} ($M=\ $Ti, V, Mn, Co) and {\xAMAB} ($A$ = Si, Ga, Ge) have also been conducted to understand and control the magnetic properties \cite{TiV, MnFefirst, MnFeferri, DuMnFe, MnFePRM, GaGe, SiGa, AlSi, AlGeGa}.
In particular, {\xMAB} is of interest because the magnetic correlation and uniaxial anisotropy are different between {\FeMAB} and {\MnMAB}. It is known that introducing antiferromagnetic correlation extends the temperature range of the magnetic transition \cite{Tad}. Moreover, coexisting ferromagnetic and antiferromagnetic correlations and competing uniaxial magnetic anisotropy along the $a$ and $b$ axes can enrich the magnetic phase diagram and yield some specific events in magnetism such as a metamagnetic transition, which can potentially broaden the application areas.
However, there are some inconsistencies among the several previous reports on the magnetic properties of {\xMAB}.
First, Chai {\etal} synthesized polycrystalline samples and performed magnetization measurements for $x \leq 0.8$. They found a gradually suppressed ferromagnetic correlation by the Mn substitution \cite{MnFefirst}.
Subsequently, Using magnetization and powder X-ray diffraction (XRD) measurements for $x \leq 0.25$ Du {\etal} proposed a re-entrant spin-glass transition below $\TC$ at around $x = 0.25$ due to chemical disorder and competing ferromagnetic and antiferromagnetic interactions \cite{DuMnFe}. In addition, Cedervall {\etal} predicted a disordered ferrimagnetic state for $x=0.5$ \cite{MnFeferri}. 
Recently, Potashnikov {\etal} synthesized polycrystalline samples for $x=0$--0.5, 0.75 and 1, and performed magnetization and neutron powder diffraction (ND) measurements to obtain a $T$--$x$ magnetic phase diagram \cite{MnFePRM}. They observed an antiferromagnetic Bragg reflection for $x \geq 0.19$ and predicted a canted antiferromagnetic phase at intermediate Mn concentration. On the other hand, the corresponding anomaly was not observed in the magnetization measurements. The XRD and ND measurements also suggested that Mn atoms are homogenously distributed in the sample, which is contrary to the previous reports \cite{MnFefirst, DuMnFe, MnFeferri}.
The inconsistencies in the magnetism of {\xMAB} arises from the difficulty in obtaining a single phase of the samples; these samples contain impurities such as \ce{Al_{13}(Fe,Mn)_4}, \ce{Al_2O_3} and \ce{(Fe,Mn)B}. Moreover, there is no report on the synthesis of {\xMAB} single crystals except for $x=0$ and 1.
The single crystal growth can unravel a new magnetic phase as well as clarify the anisotropic magnetic properties, which could be helpful for improving the magnetocaloric property of {\FeMAB} by tuning the magnetocrystalline anisotropy. \\ 
\indent In this paper, we report the synthesis of {\xMAB} single crystalline samples and discuss their magnetic properties based on magnetization measurements. We newly found a magnetic transition at intermediate Mn concentration, and then constructed detailed $T$--$x$, $H$--$T$, and three-dimensional $H$--$T$--$x$ magnetic phase diagrams.

\section{Experiments}
\subsection{Crystal growth}
Single crystals of {\xMAB} were grown by the Al self-flux method described in literature \cite{FeSCMAB, MnSCMAB, CrSCMAB}. Al shots (Rare Metallic, 99.999\%), Fe powder (Rare Metallic, 99.9\%), crushed Mn flakes (Rare Metallic, 99.9\%), and B chunks (Furuuchi Chemical, 99\%) were used. Al shots and mixture of Fe, Mn, and B powders were placed in a boron nitride crucible and sealed in a quartz tube with a partial pressure of argon gas.\\
\indent Table\ \ref{syn} shows the initial compositions of the raw materials and the nominal Mn concentrations $x$, which is the molar ratio of $\ce{Mn}/(\ce{Fe}+\ce{Mn})$. In the case of nominally $0 \leq x \leq 0.3$, single crystals of {\xMAB} were successfully grown by adopting an initial Al/(Fe+Mn) molar ratio of less than 3, which otherwise resulted in forming only \ce{Al_13(Fe,Mn)_4} single crystals when nominally $x=0$--0.17. The ampoule was initially heated up to 1200 {\degC} over 2\ h and held for 3\ h, and then cooled down to 1180 {\degC} over 1 h to avoid peritectic reactions, and slowly cooled down to 1080 {\degC} over 30 h, at which the samples were centrifuged to separate single crystals from the flux. Plate-like crystals similar to those of {\FeMAB} \cite{FeSCMAB} were obtained. The size was up to $5\times5\times 0.2\ {\rm mm}^3$ in the range of nominal $x=0$--0.17 (actual $x=0$--0.36), while up to $1\times1\times 0.1 \ {\rm mm}^3$ for nominal $x = 0.3$ (actual $x=0.46$) (Fig.\ \ref{str}(d)).  
On the other hand, a large Al/(Mn+Fe) molar ratio was applicable for crystal growth with nominal $x \geq 0.3$. In this case, the tube was initially heated up to 1200 {\degC} over 2\ h, held for 3\ h, and slowly cooled down to 800$-$900 {\degC} at a rate of 5 {\degC}/h. Using a centrifuge, we separated strip-shaped crystals (Fig.\ \ref{str}(e)) similar to those of {\MnMAB} \cite{MnSCMAB}. The size was up to $3\times1\times0.1\ {\rm mm}^3$ when $\ce{Al}/(\ce{Fe}+\ce{Mn}) \leq 5$, while up to $10\times1\times0.2\ {\rm mm}^3$ when $\ce{Al}/(\ce{Fe}+\ce{Mn}) \geq 10$. Hexagonal plate-like crystals of \ce{AlB_2} and prismatic crystals of \ce{Al_13(Fe,Mn)_4} were also obtained from the melts with initial compositions of \ce{Al_5Fe_{0.7}Mn_{0.3}B} and \ce{Al_5Fe_{0.5}Mn_{0.5}B}. Note that the actual Mn concentration of {\xMAB} depends significantly on the Al/(Mn+Fe) ratio, as described below. The flux and impurity phases \ce{Al_13(Fe,Mn)_4} and \ce{AlB_2} remaining on the surface were removed by dilute HClaq immersion and concentrated KOHaq etching.

\subsection{Crystal characterization and magnetic measurements}\label{expchr}
The grown crystals were characterized by powder X-ray diffraction (XRD) measurements with Cu\ $K_{\alpha_1}$ lines using X’Pert PRO Alpha-1 (PANalytical). The Rietveld refinement was performed using Rietan-FP \cite{FP}. Powder samples were prepared by crushing parts of the single crystals.
The chemical composition of the grown crystals was analyzed by wave length dispersive X-ray spectroscopy (WDX). We also measured the compositions of several batches by inductively coupled plasma mass spectrometry (ICP-MS) to estimate the content of boron, which was difficult to measure accurately by WDX. 
Crystal axes were identified using a Laue camera. Magnetization measurements were performed using a SQUID magnetometer (MPMS, Quantum Design) in the temperature range of 5--700 K and under magnetic fields up to 7 T. The magnetization above room temperature was measured using an oven option of the MPMS.
For a temperature range of $T=5$--350 K, a piece of single crystal was used for magnetization measurement when $x =0$--0.46, while more than five crystals were used when $x =0.65$--1 with small crystal size and magnetization. 
For the high-temperature measurements using the oven option, we used several other single crystals in the same batch.
\begin{table*}[ht]
  \begin{ruledtabular}
  \caption{Initial compositions and actual compositions determined by WDX and ICP-MS. $x$ corresponds to Mn/(Mn+Fe).} 
 \begin{center}
  \begin{tabular}{lccccc}
  &&\multicolumn{2}{c}{WDX}&\multicolumn{2}{c}{ICP-MS}\\
  \cmidrule(lr){3-4} \cmidrule(lr){5-6}
  Initial composition & Nominal $x$ & Composition & $x$ & Composition & $x$ \\
  \hline
  \ce{Al_5Fe_2B_{1.33}} & 0 & \ce{Fe_2Al_{1.00}B_{(-)}}  & 0  & \ce{Fe_2Al_{1.05}B_{1.93}} & 0 \\
  \ce{Al_5Fe_2Mn_{0.1}B_{1.4}} & 0.05 & \ce{(Fe_{0.87}Mn_{0.13})_2Al_{0.96}B_{(-)}} & 0.13  & - & - \\
  \ce{Al_5Fe_2Mn_{0.2}B_{1.47}} & 0.09 & \ce{(Fe_{0.79}Mn_{0.21})_2Al_{0.99}B_{(-)}}  & 0.21  & \ce{(Fe_{0.79}Mn_{0.21})_2Al_{1.04}B_{2.00}} & 0.21 \\
  \ce{Al_5Fe_2Mn_{0.4}B_{1.6}} & 0.17 & \ce{(Fe_{0.69}Mn_{0.31})_2Al_{0.98}B_{(-)}}  & 0.31  & \ce{(Fe_{0.68}Mn_{0.32})_2Al_{0.98}B_{2.00}} & 0.32 \\
  \ce{Al_5Fe_{1.68}Mn_{0.32}B_{1.6}} & 0.16 & \ce{(Fe_{0.64}Mn_{0.36})_2Al_{0.98}B_{(-)}}  & 0.36  & \ce{(Fe_{0.64}Mn_{0.36})_2Al_{1.09}B_{2.00}} & 0.36 \\
  \ce{Al_{2.08}Fe_{0.7}Mn_{0.3}B_{0.83}} & 0.3 & \ce{(Fe_{0.54}Mn_{0.46})_2Al_{0.93}B_{(-)}}  & 0.46  &  \ce{(Fe_{0.52}Mn_{0.48})_2Al_{1.02}B_{1.99}} & 0.48 \\
  \ce{Al_{3.5}Fe_{0.7}Mn_{0.3}B} & 0.3 & \ce{(Fe_{0.47}Mn_{0.53})_2Al_{0.97}B_{(-)}}  & 0.53  & \ce{(Fe_{0.44}Mn_{0.56})_2Al_{1.01}B_{2.01}} & 0.56 \\
  \ce{Al_5Fe_{0.7}Mn_{0.3}B} & 0.3 & \ce{(Fe_{0.36}Mn_{0.65})_2Al_{0.95}B_{(-)}}  & 0.65  & \ce{(Fe_{0.34}Mn_{0.66})_2Al_{1.01}B_{2.06}} & 0.66 \\
  \ce{Al_{3.5}Fe_{0.5}Mn_{0.5}B} & 0.5 & \ce{(Fe_{0.26}Mn_{0.74})_2Al_{0.96}B_{(-)}}  & 0.74  & - & - \\
  \ce{Al_{10}Fe_{0.5}Mn_{0.5}B} & 0.5 & \ce{(Fe_{0.13}Mn_{0.87})_2Al_{0.98}B_{(-)}}  & 0.87  & \ce{(Fe_{0.13}Mn_{0.87})_2Al_{1.05}B_{1.99}} & 0.87 \\
  \ce{Al_{20}MnB} & 1 & \ce{Mn_2Al_{0.99}B_{(-)}}  & 1  & \ce{Mn_2Al_{1.06}B_{1.94}} & 1 \\
  \end{tabular}
  \label{syn}
  \end{center}
  \end{ruledtabular}
\end{table*}
\section{Results}
\subsection{Sample characterization}
\begin{figure}[ht]
    \centering
    \includegraphics[keepaspectratio, width=1\columnwidth]{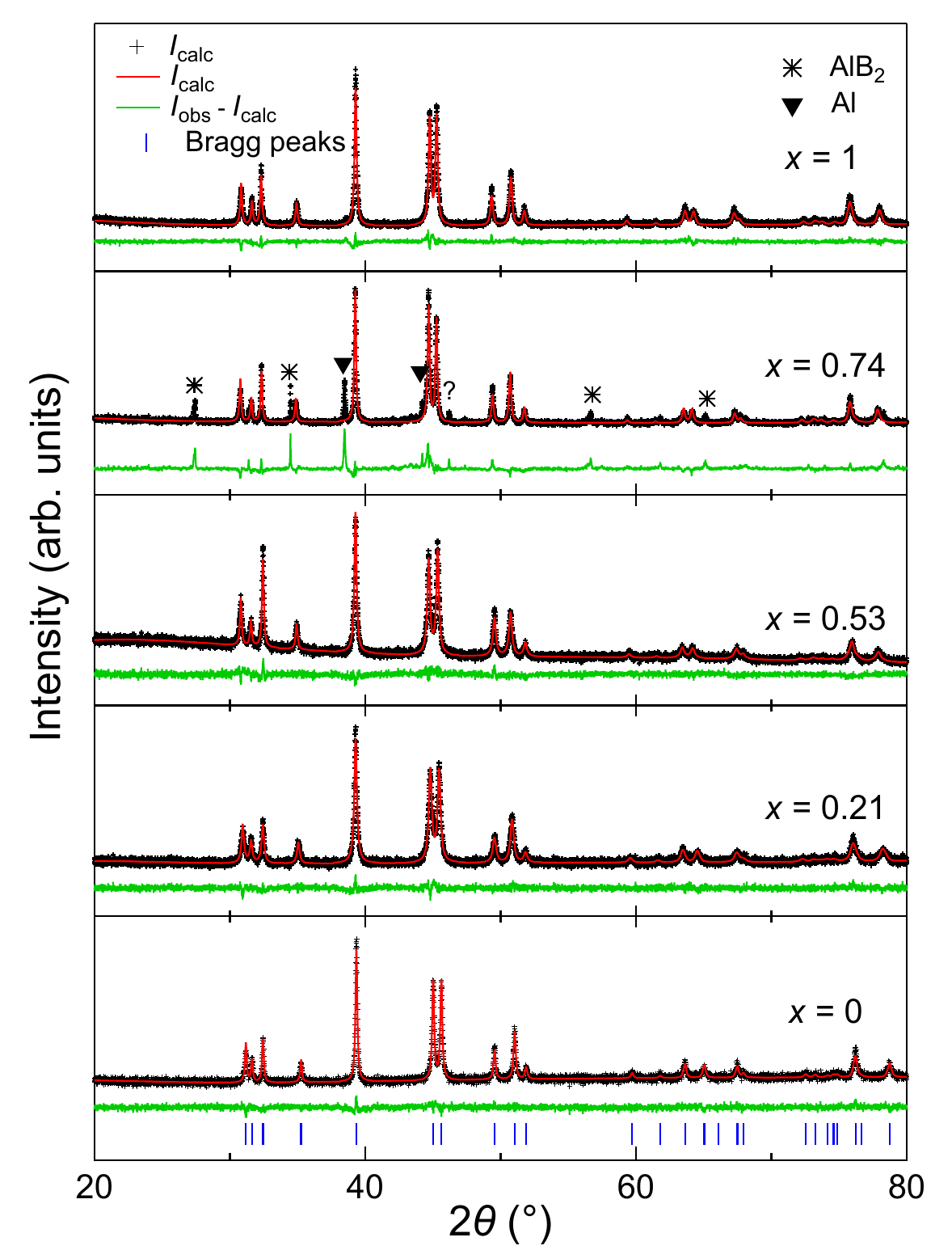}
    \caption{Powder X-ray diffraction profiles of  {\xMAB} at room temperature. Results of the Rietveld refinement and expected Bragg reflections are also shown.}
    \label{xrd}
\end{figure}
Table\ \ref{syn} shows the chemical compositions of the grown crystals estimated by WDX. These were measured and averaged for more than three spots on the crystal surface and determined as averages for more than three crystals in the same batch. We have confirmed that the compositions of different batches prepared under the same conditions are practically identical. The listed compositions are normalized to $\ce{Fe}+\ce{Mn}=2$ and the actual Mn concentration $x$ is equal to Mn/(Mn+Fe).
Table\ \ref{syn} shows single crystals of {\xMAB} have been successfully synthesized in the entire composition range. A slight deviation from the stoichiometry in Al is attributed to the mutual substitution of Fe and Al for the respective lattice sites, which has also been reported for {\FeMAB} \cite{nonsto}. Actual $x$ is higher than nominal $x$ and depends on the initial Al/(Mn+Fe) molar ratio. For example, actual $x$ obviously increased with initial Al/(Mn+Fe), despite the same nominal $x=0.5$. 
We note that the standard deviations of $x$ for all measured crystals were less than 0.05, suggesting no segregation into Fe-rich and Mn-rich aggregates in the crystal, consistent with the result of the polycrystalline samples reported in Ref.\ \cite{MnFePRM} rather than Ref.\ \cite{DuMnFe}. \\
\indent As described in section \ref{expchr}, it was difficult to determine accurately the boron concentration by WDX. We performed ICP-MS and found that the compositions are in accordance with the ideal compotision ratio of $({\rm Fe} + {\rm Mn}):{\rm Al}:{\rm B} = 2:1:2$ and the actual $x$ values agree well with those estimated by WDX (see Table\ \ref{syn}). The overestimation of Al concentration may be due to residual flux at the surface. We will use the values measured by WDX as the Mn concentration in th following discussion.\\
\indent Using the Laue method, the crystal orientation perpendicular to the flat surface was found to be [010] for all Mn concentrations. For the strip-shaped crystals, the longitudinal and transverse directions were identified as [100] and [001], respectively. These results are in agreement with previous reports of {\FeMAB} and {\MnMAB} \cite{FeSCMAB, MnSCMAB}.\\
\begin{table*}[ht]
  \begin{ruledtabular}
  \caption{Refined lattice parameters, volume ($V$), atomic coordinates $y(4j)$ in Fe $4j\ (0,y,0)$ site and $y(4i)$ in B $4i\ (0,y,0)$, the reliability $R$ factors ($R_{\rm wp}$ and $R_{\rm p}$), and the goodness-of-fit indicator $S$. Literature values for non-doped compounds are also shown.} 
\begin{center}
  \begin{tabular}{lccccccccc}
  $x$ & $a$\ (\AA) & $b$\ (\AA) & $c$\ (\AA) & $V$\ (\AA$^3$) & $y(4j)$ & $y(4i)$ & $R_{\rm{wp}}$ & $R_{\rm{p}}$ & $S$ \\
  \hline
  0 & 2.9227(1) & 11.0268(4) & 2.8653(1) & 92.34  & 0.3544(1) & 0.2075(7) & 3.814 & 3.035 & 1.1019 \\
  0.21 & 2.9295(1) & 11.0306(4) & 2.8862(1) & 93.27  & 0.3542(1) & 0.2098(7) & 4.069 & 3.244 & 1.0626 \\
  0.31 & 2.9307(1) & 11.0288(4) & 2.8939(1) & 93.54  & 0.3542(1) & 0.2088(7) & 4.572 & 3.623 & 1.0695 \\
  0.36 & 2.9316(1) & 11.0278(4) & 2.8947(1) & 93.58  & 0.3543(1) & 0.2082(7) & 4.419 & 3.518 & 1.0809 \\
  0.46 & 2.9302(1) & 11.0235(5) & 2.8979(1) & 93.61  & 0.355(1) & 0.2006(6) & 4.539 & 3.562 & 1.1091 \\
  0.53 & 2.9306(1) & 11.0289(4) & 2.8994(1) & 93.71  & 0.3555(1) & 0.1972(7) & 6.11 & 4.84 & 1.094 \\
  0.65 & 2.9295(1) & 11.0404(3) & 2.9013(1) & 93.84  & 0.3547(1) & 0.2048(6) & 5.509 & 4.253 & 1.2322 \\
  0.74 & 2.9292(2) & 11.0452(7) & 2.9026(2) & 93.91  & 0.3595(2) & 0.1839(14) & 12.949 & 8.247 & 2.5961 \\
  0.87 & 2.9265(1) & 11.0591(3) & 2.9015(1) & 93.90  & 0.3551(1) & 0.2075(6) & 6.206 & 4.951 & 1.0941 \\
  1 & 2.9225(1) & 11.0720(3) & 2.8968(1) & 93.74  & 0.3554(1) & 0.2065(5) & 6.583 & 5.229 & 1.1537 \\
  \hline
  0 \cite{FeSCMAB} & 2.9617  & 11.0330  & 2.8660  & 92.23  & 0.3539  & 0.2066  & -- & -- & -- \\
  0 \cite{MnFePRM} & 2.9261  & 11.0316  & 2.8677  & 92.57  &  -- & -- & -- & -- & -- \\
  1 \cite{MnSCMAB} & 2.9215  & 11.0709  & 2.8972  & 93.71  & 0.3552  & 0.2065  & -- & -- & -- \\
  1 \cite{MnFePRM} & 2.9202  & 11.0613  & 2.8957  & 93.64  & -- & -- & -- & -- & -- \\
  \end{tabular}
  \label{riet}
  \end{center}
  \end{ruledtabular}
\end{table*}
\begin{figure}[ht]
    \begin{minipage}[b]{0.49\linewidth}
      \centering
      \includegraphics[keepaspectratio, width=1\columnwidth]{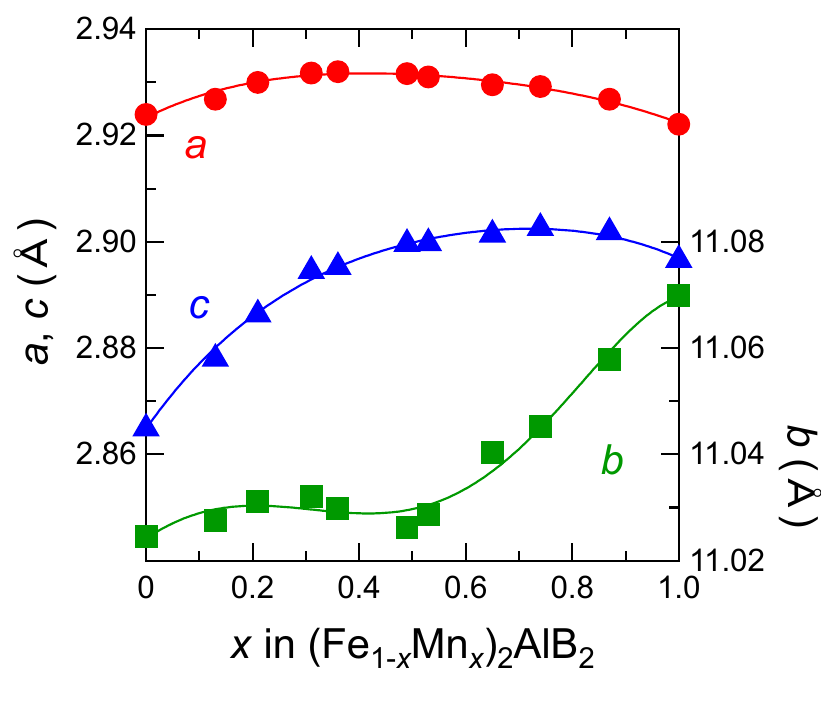}
    \end{minipage}
    \begin{minipage}[b]{0.49\linewidth}
      \centering
      \includegraphics[keepaspectratio, width=1\columnwidth]{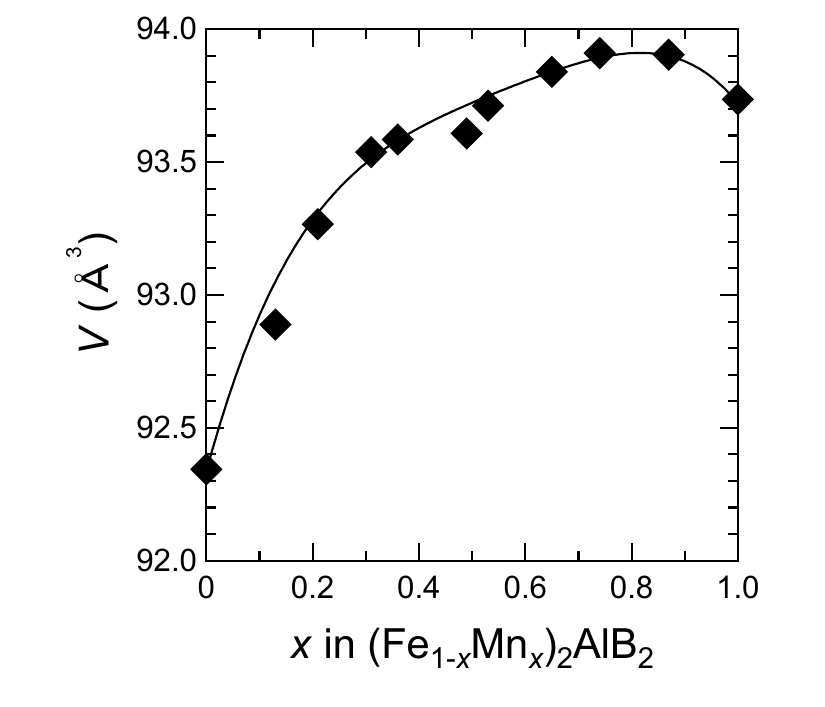}
    \end{minipage}
    \caption{Composition dependences of (a) lattice parameters and (b) unit-cell volume for {\xMAB}. The Mn concentration $x$ was determined by WDX analysis.}
    \label{LP}
\end{figure}
\indent Figure\ \ref{xrd} shows the powder XRD profiles of {\xMAB} with selected Mn concentrations at room temperature. We used powder samples prepared by crushing pieces of the single crystals. The reflections of the main phase for all Mn concentrations are indexed by the planes of an orthorhombic phase with the space group $Cmmm$. A small amount of Al and \ce{AlB_2} appear as residual flux and a by-product on the crystal surface, respectively, in the XRD patterns of $x=0.65$ and 0.74 because of insufficient etching.\\
\indent The XRD patterns were refined by the Rietveld method using Rietan-FP \cite{FP}. Due to the cleavage tendency, the \{010\}-preferred orientation was employed in the refinement. 
It was difficult to directly distinguish between the Fe and Mn atoms due to the similar atomic scattering factors, and optimizing all parameters at once resulted in negative site occupancies and isotropic displacement parameters $\Biso$, which are physically unreasonable. 
To avoid this problem, all $\Biso$ and site occupancies were fixed and all the other parameters were refined; we assumed perfect ordering of Al and B atoms at the 2$a$ and 4$i$ sites, respectively, and used values estimated by WDX for the site occupancies of Mn and Fe at the $4j$ site and literature values for $\Biso$ ($\Biso^{\rm Al} =\Biso^{\rm Mn}=\Biso^{\rm Fe}=0.5$ and $\Biso^{\rm B} = 0.7$ \cite{FeSCMAB, MnSCMAB}). 
The results of the fitting are shown in Fig.\ \ref{xrd} and Table\ \ref{riet}. Except for the samples with $x=0.65$ and 0.74, the small $R$-factors and the goodness-of-fit indicator $S$ demonstrate satisfactory refinements. The unit-cell volumes of {\FeMAB} and {\MnMAB} are calculated to be 92.34 {\AA$^3$} and 93.74 {\AA$^3$}, respectively, which are close to the reported values of single crystals (92.23 {\AA$^3$} \cite{FeSCMAB} and 93.71 {\AA$^3$} \cite{MnSCMAB}).
Figure\ \ref{LP} shows the refined lattice parameters and unit-cell volumes of {\xMAB} as a function of  Mn concentration $x$. Our results are in good agreement with but more detailed than the previous reports of polycrystalline samples \cite{MnFePRM, DuMnFe}. These values do not follow Vegard's law; the lattice parameters $a$ and $c$ start to decrease gradually at $x\simeq 0.5$ and 0.8, respectively, whereas $b$ has a broad hump at around $x=0.25$, which was missing in the previous reports \cite{MnFePRM, DuMnFe}, and increases above $x \simeq0.5$. The unit-cell volume shrinks above $x \simeq 0.8$ after expanding with increasing $x$. 
This nonlinear and nonmonotonic variation suggests that the spontaneous magnetovolume effect at room temperature depends sensitively on the Mn concentration $x$ because the magnetic transition temperature varies at around room temperature, as shown in Fig.\ \ref{Tx}.
\subsection{Magnetization}
\begin{figure*}[t]
\begin{tabular}{ccc}
  \begin{minipage}[h]{0.3\linewidth}
    \centering
    \includegraphics[keepaspectratio, width=\columnwidth]{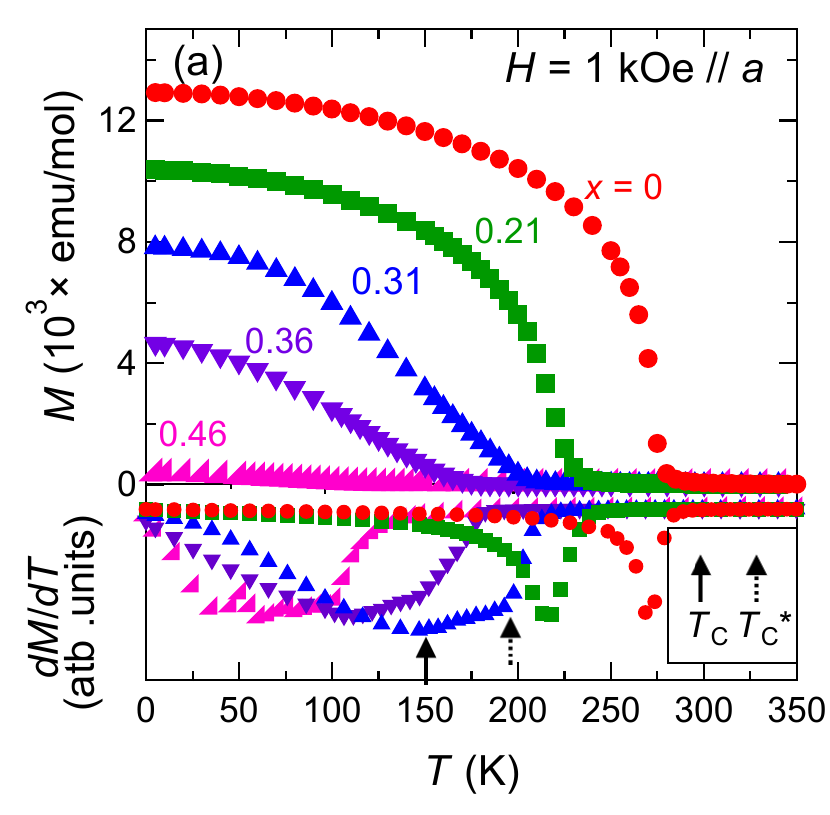}
  \end{minipage}&
  \begin{minipage}[h]{0.3\linewidth}
    \centering
    \includegraphics[keepaspectratio, width=\columnwidth]{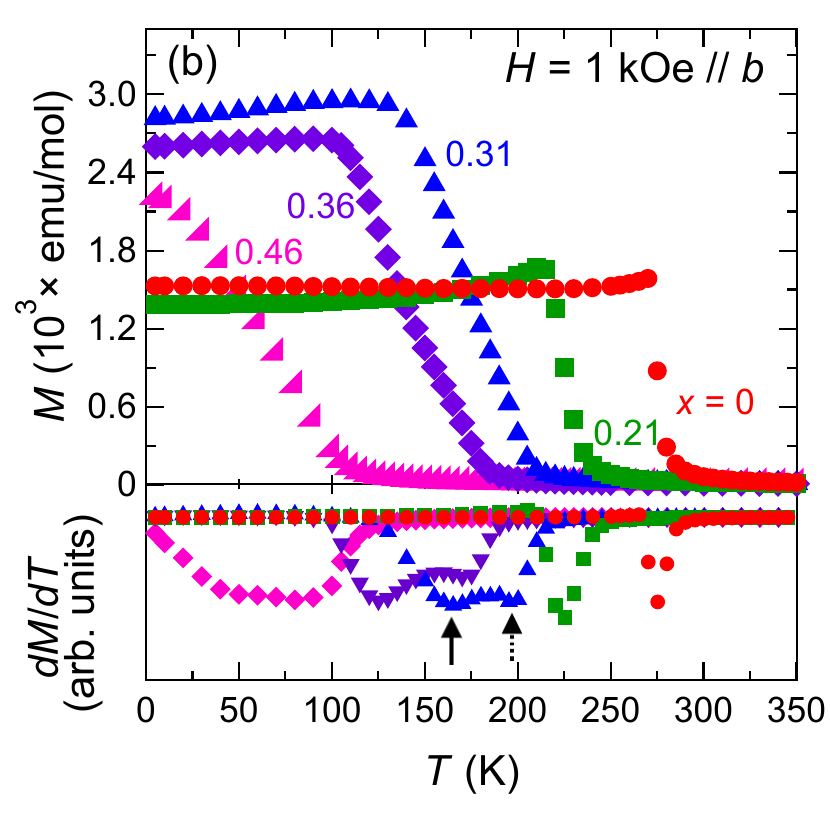}
  \end{minipage}&
  \begin{minipage}[h]{0.3\linewidth}
    \centering
    \includegraphics[keepaspectratio, width=\columnwidth]{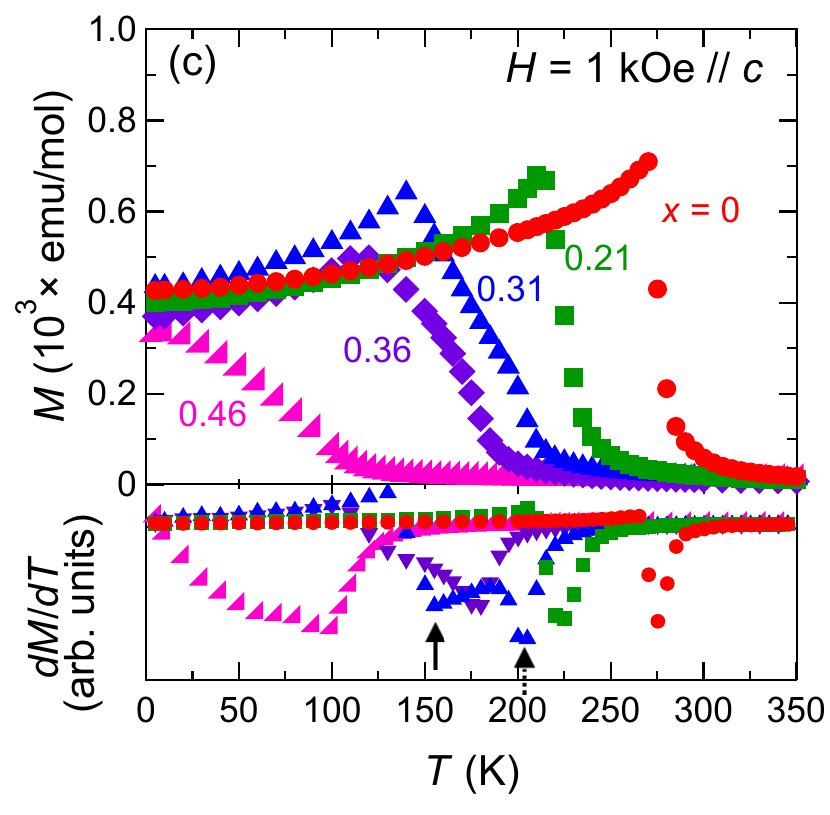}
  \end{minipage}\\
    \begin{minipage}[h]{0.3\linewidth}
    \centering
    \includegraphics[keepaspectratio, width=\columnwidth]{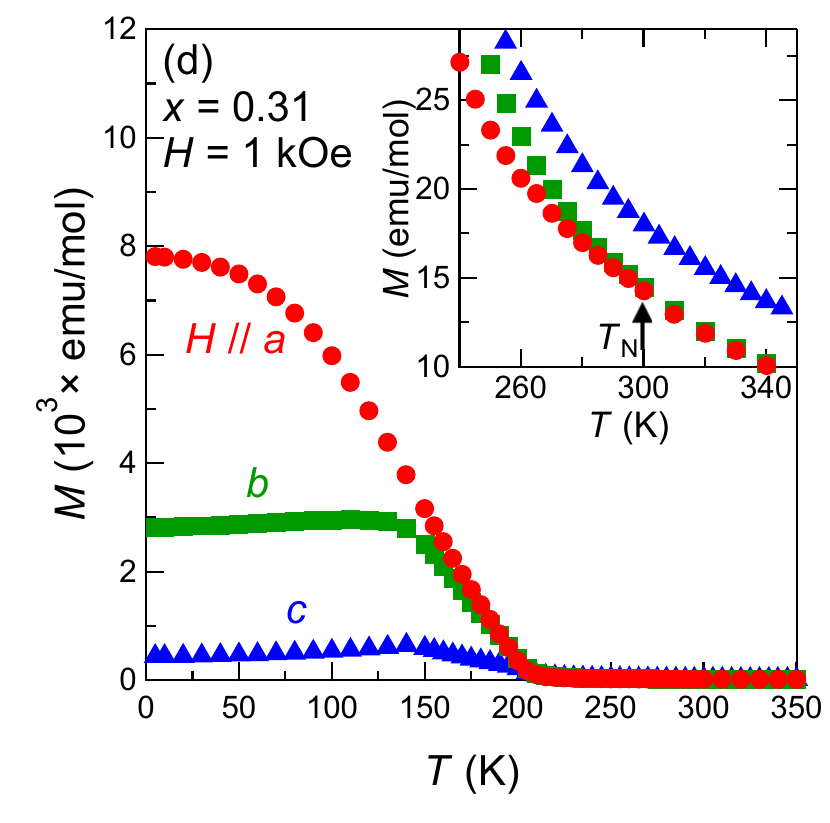}
  \end{minipage}&
  \begin{minipage}[h]{0.3\linewidth}
    \centering
    \includegraphics[keepaspectratio, width=\columnwidth]{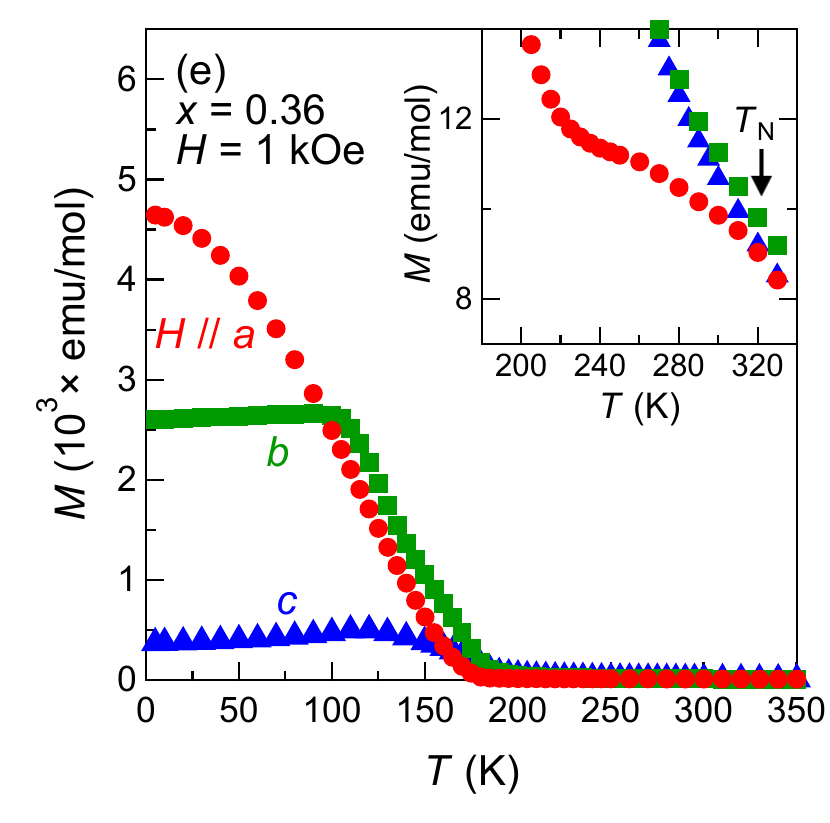}
  \end{minipage}&
  \begin{minipage}[h]{0.3\linewidth}
    \centering
    \includegraphics[keepaspectratio, width=\columnwidth]{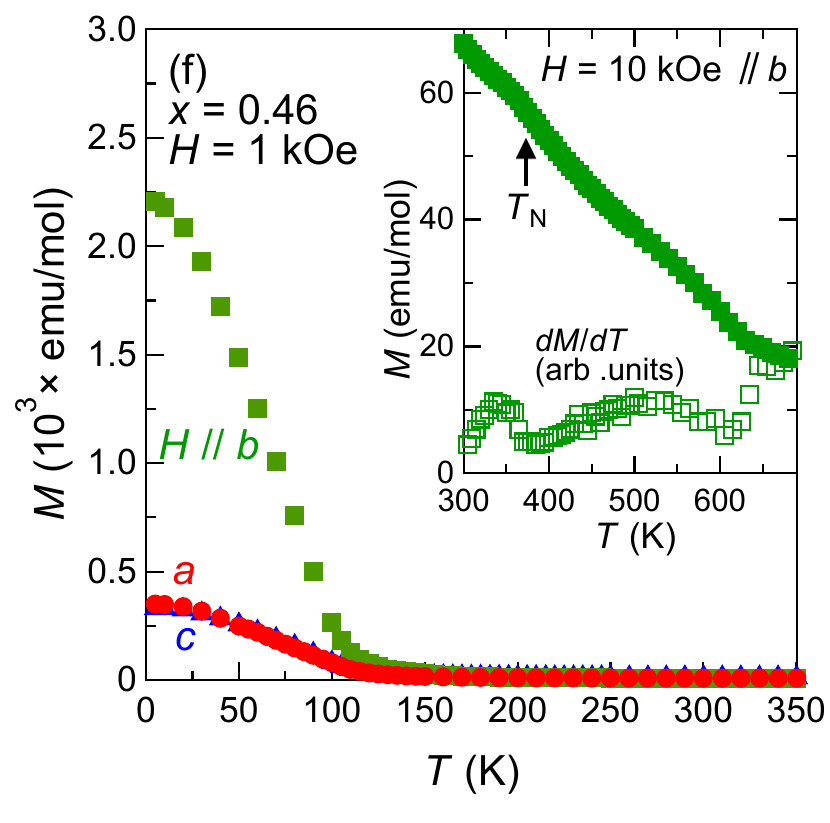}
  \end{minipage}
\end{tabular}
  \caption{(a)--(c) Temperature dependences of the magnetization and its derivative $dM/dT$ for $x=0$--0.46 under a magnetic field of $H=1$ kOe along the $a$, $b$, and $c$ axes. The arrow and dotted arrow show $\TC$ and $\TCC$, respectively. (d)--(f) Temperature dependences of the magnetization along the three crystallographic axes for $x=0.31$, 0.36 and 0.46. The insets show the data in high-temperature regions.}
  \label{MTlowx}
\end{figure*}
\begin{figure}[t]
    \centering
    \includegraphics[keepaspectratio, width=0.7\columnwidth]{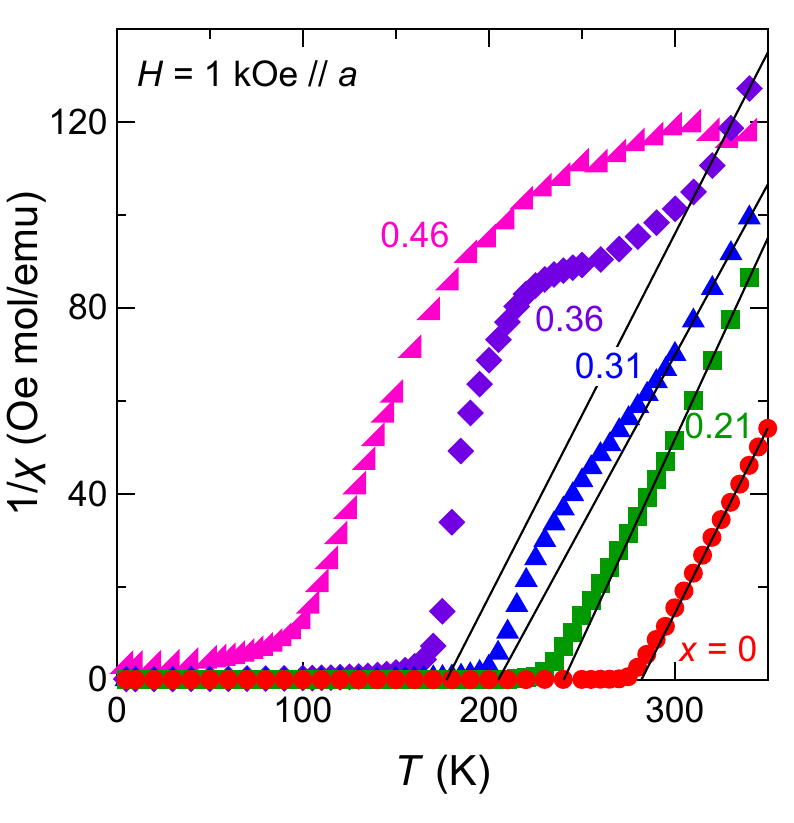}
    \caption{Temperature dependences of the inverse susceptibility for $x=0$--0.46 along the $a$ axis. The lines represent the fit to the Curie-Weiss law.}
    \label{inv}
\end{figure}
\begin{figure*}[t]
\begin{tabular}{cccc}
  \begin{minipage}[h]{0.28\linewidth}
    \centering
    \includegraphics[keepaspectratio, width=\columnwidth]{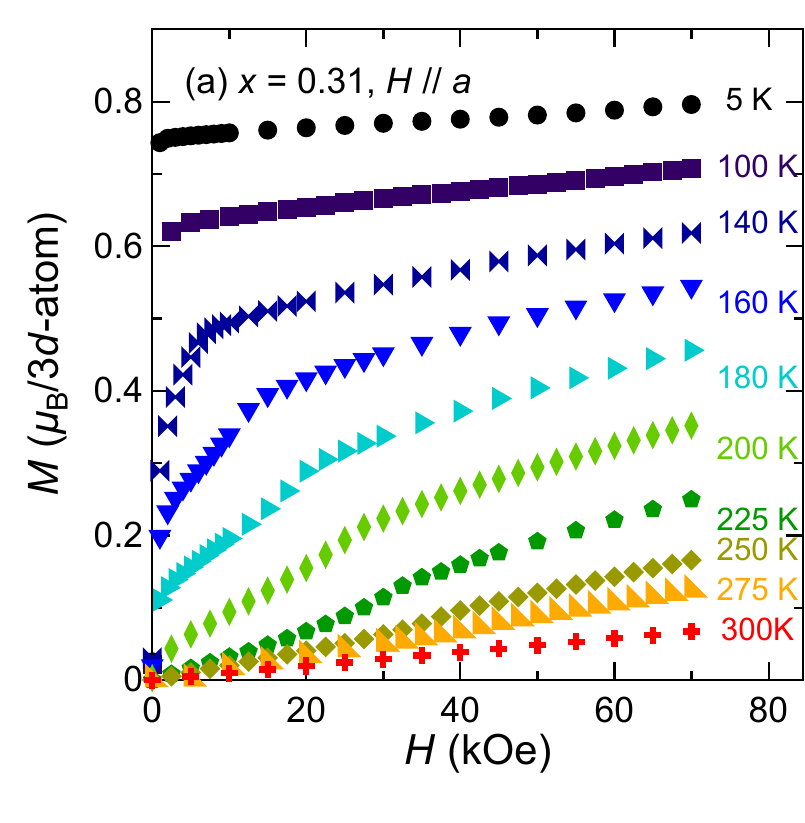}
  \end{minipage}&
    \begin{minipage}[h]{0.16258\linewidth}
    \centering
    \includegraphics[keepaspectratio, width=\columnwidth]{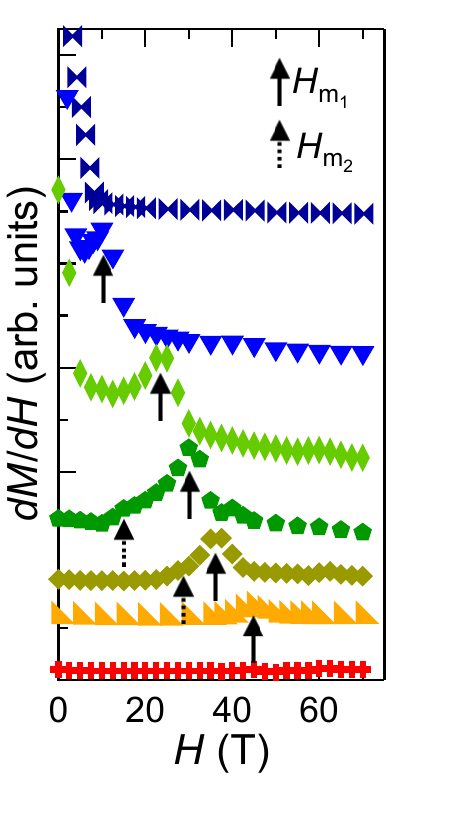}
  \end{minipage}&
    \begin{minipage}[h]{0.28\linewidth}
    \centering
    \includegraphics[keepaspectratio, width=\columnwidth]{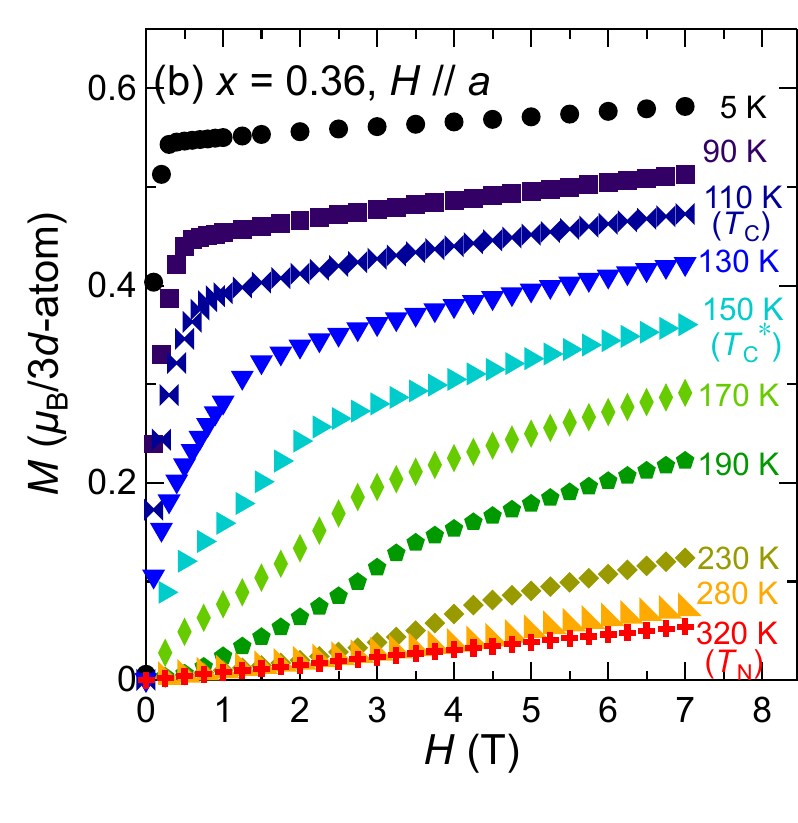}
  \end{minipage}&
    \begin{minipage}[h]{0.16258\linewidth}
    \centering
    \includegraphics[keepaspectratio, width=\columnwidth]{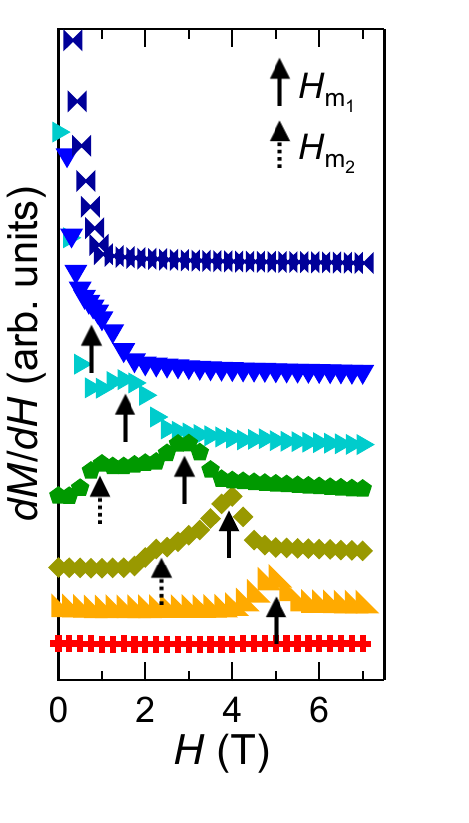}
  \end{minipage}\\
    \begin{minipage}[h]{0.28\linewidth}
    \centering
    \includegraphics[keepaspectratio, width=\columnwidth]{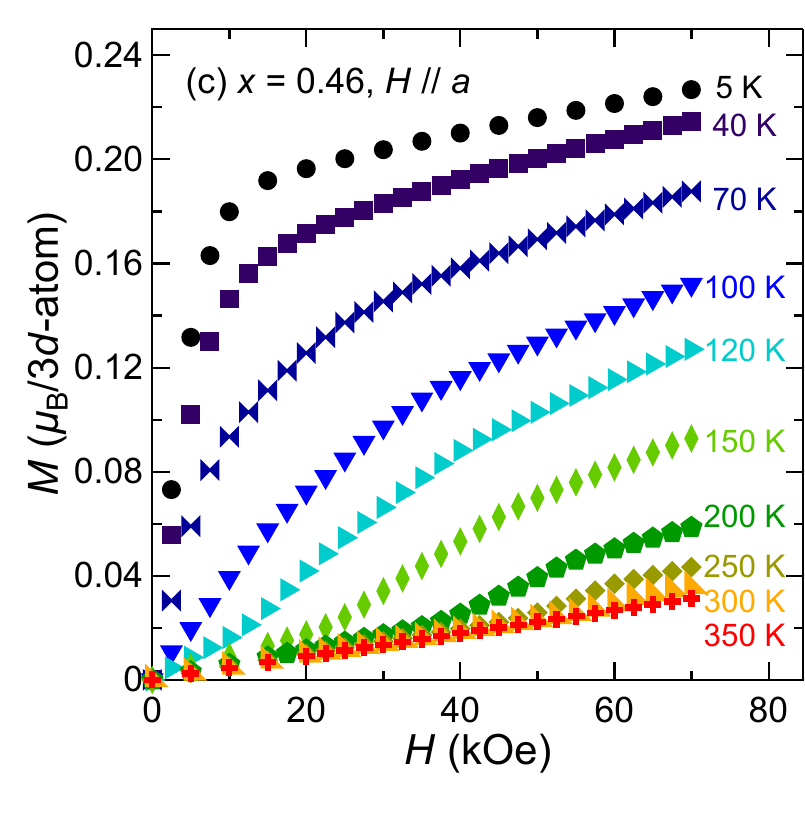}
  \end{minipage}&
    \begin{minipage}[h]{0.16258\linewidth}
    \centering
    \includegraphics[keepaspectratio, width=\columnwidth]{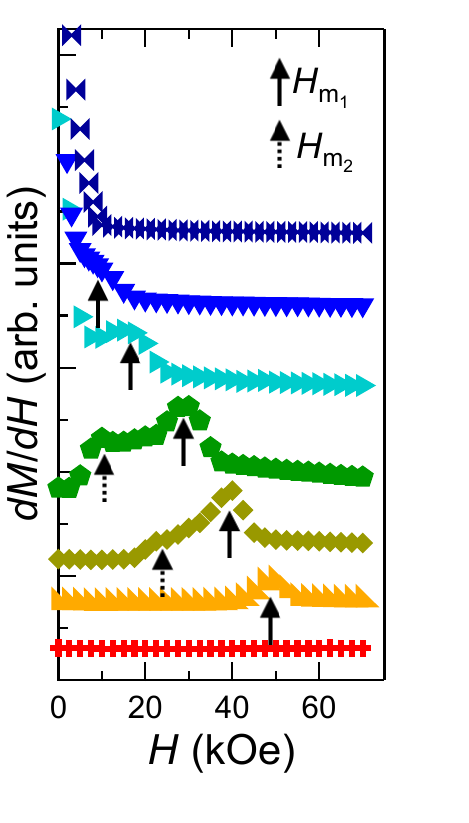}
  \end{minipage}&
    \begin{minipage}[h]{0.28\linewidth}
    \centering
    \includegraphics[keepaspectratio, width=\columnwidth]{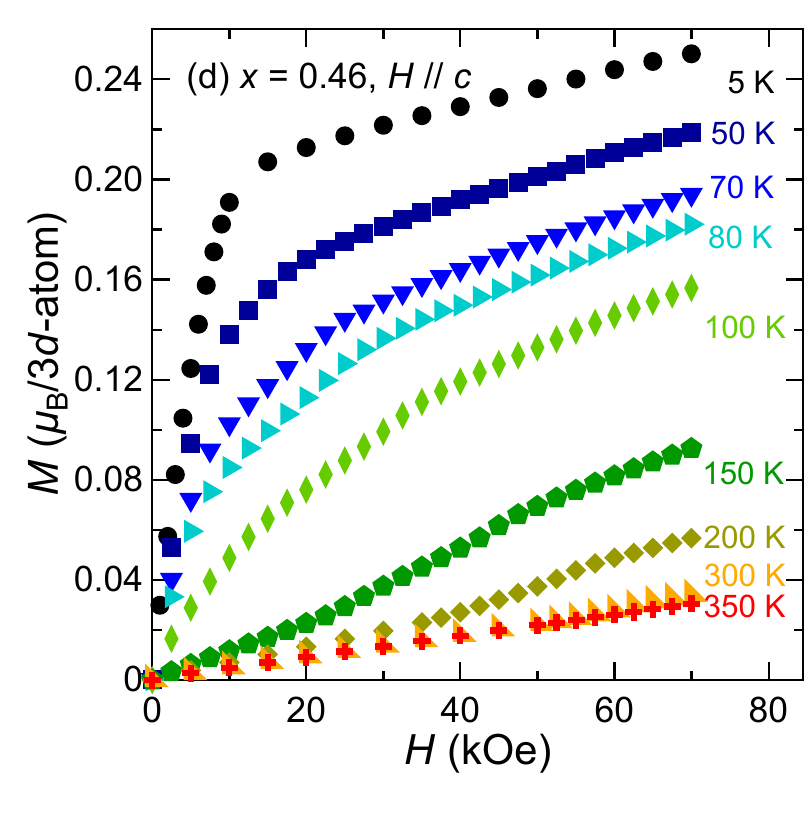}
  \end{minipage}&
    \begin{minipage}[h]{0.16258\linewidth}
    \centering
    \includegraphics[keepaspectratio, width=\columnwidth]{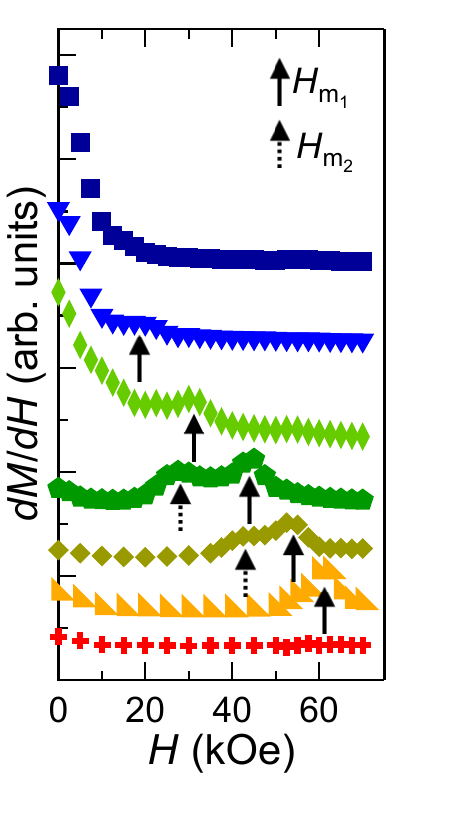}
  \end{minipage}
\end{tabular}
  \caption{Field dependences of the magnetization measured at different temperatures parallel to the $a$ axis for $x=0.31$ (a),  0.36 (b), and 0.46 (c), and parallel to the $c$ axis for $x=0.46$ (d).}
  \label{metalowx}
\end{figure*}

\begin{table*}[t]
\begin{ruledtabular}
\caption{Transition temperatures ($\TC, \TCC, \TN$) and the results of Curie-Weiss fitting ({$\peff, \TCW$}) for {\xMAB}. $\TC, \TCC, \peff$, and $\TCW$ were estimated from the magnetization along the $a$ axis.} 
\begin{center}
\begin{tabular}{lcccccc}
$x$ (K) & $\TC$ (K) & $\TCC$ (K) & $\TN$ (K) & $\peff$ ($\uB$/3$d$-atom) & $\TCW$ (K) & $\Msat$ ($\uB$/3$d$-atom) \\
\hline
0 & 275(5) & -- & -- & 2.28 & 280 & 1.22 \\
0.21 & 220(5) & -- & -- & 2.15 & 240 & 0.93 \\
0.31 & 150(10) & 205(10) & 300(20) & 2.33 & 205 & 0.71 \\
0.36 & 110(10) & 150(10) & 320(10) & 2.26 & 177 & 0.57 \\
0.46 & 40(20) & 95(20) & 350(15) & -- & -- & 0.23 \\
0.65 & -- & -- & 385(5) & -- & -- & -- \\
0.74 & -- & -- & 350(10) & -- & -- & -- \\
0.87 & -- & -- & 340(5) & -- & -- & -- \\
1 & -- & -- & 315(5) & -- & -- & -- \\
\end{tabular}
\label{transition}
\end{center}
\end{ruledtabular}
\end{table*}
Figure\ \ref{MTlowx} shows the temperature dependences of the magnetization for $x$ = 0--0.46. 
These samples show ferromagnetic behavior and the Curie temperature $\TC$ decreases with increasing Mn concentrations $x$. The Curie temperature of {\FeMAB} ($x=0$), which is $\TC = 275$ K, is close to literature data ($\TC=273$\ K, \cite{FeSCMAB}). There is neither cusp nor distinctive bifurcation between the data under zero-field cooling and field cooling at low temperature, suggesting the absence of the re-entrant spin-glass transition proposed in Ref.\ \cite{DuMnFe}. 
For $x=0.31$--0.46, a gradual ferromagnetic upturn was observed and $dM/dT$ shows two-step anomalies at $\TC$ and $\TCC$. The transition temperatures depend on the direction of the applied magnetic fields, except for $x=0$ and 0.21. 
Above $\TCC$, another magnetic transition was observed for $x=0.31$--0.46; for $x =0.31$ and 0.36 a bifurcation of the magnetization between the $a$ axis and the $b$ and $c$ axes was observed below $\TN \simeq 300$ K and 320 K, respectively (insets of Figs.\ \ref{MTlowx}(d),(e)), suggesting the presence of an antiferromagnetic ordering with spins aligned along the $a$ axis. For $x=0.46$, we can see in Fig.\ \ref{MTlowx}(f) a kink in the magnetization along the $b$ axis around 360 K, also suggesting an antiferromagnetic ordering. 
Note that another kink was observed around $T=620$ K, which is attributed to a ferromagnetic transition of impurities contained in the sample; the anomaly was absent when a piece of crystal was used (see Figs.\ \ref{metalowx}(c),(d)).\\  
\indent Figure\ \ref{inv} shows the inverse susceptibility $1/\chi$ for $x$ = 0--0.46 under the field applied along the $a$ axis. The temperature dependence of $1/\chi$ exhibits the Curie-Weiss-like paramagnetic behavior above $\TC$ in the cases of $x=0$ and 0.21, while shows a hump-anomaly in $x$ = 0.31--0.46. The hump is considered to appear at $\TN$, supporting the presence of the antiferromagnetic phase suggested from Figs.\ \ref{MTlowx}(d)--(f). 
The susceptibility in the paramagnetic region for $x$ = 0--0.36 was fitted by the Curie-Weiss law $\chi = C/(T-\TCW)$, where $C$ is the Curie constant and $\TCW$ the Curie-Weiss temperature. The estimated parameters are shown in Table\ \ref{transition}. The values of $\TCW$ are all positive, and at $x=0.31$ and 0.36 they are close to $\TCC$. The effective moment $\peff$, estimated from $C$ using the relation $ C = \uB \peff ^2/3\kB$, does not vary significantly with $x$. \\
\begin{figure*}[t]
    \centering
    \includegraphics[keepaspectratio, width=1.8\columnwidth]{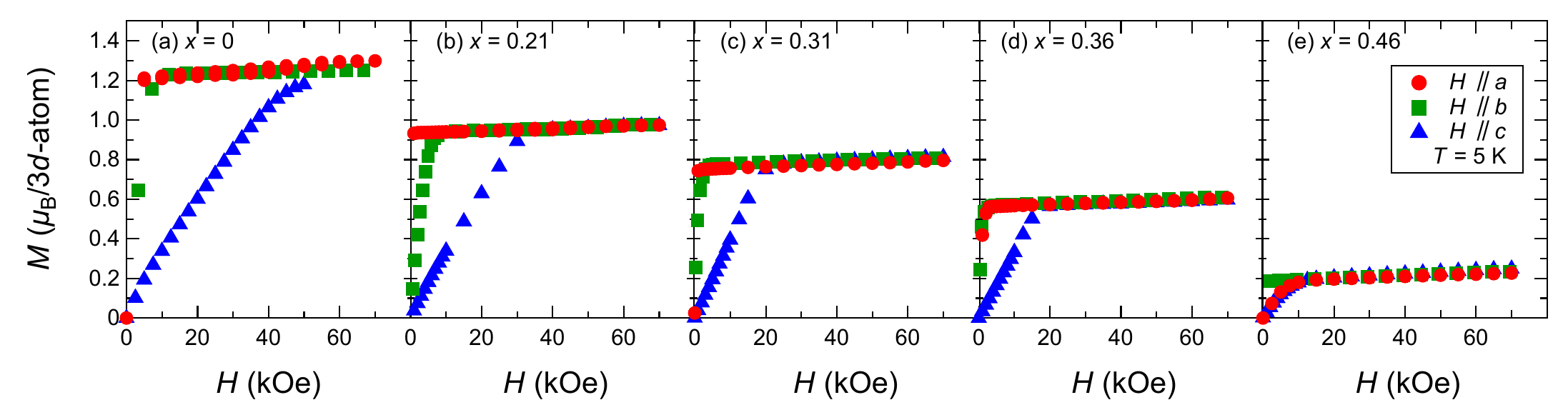}
    \caption{Field dependences of the magnetization for $x=0$--0.46 at $T=5$ K.}
    \label{MHlowx}
\end{figure*}
\begin{figure}[t]
    \centering
    \includegraphics[keepaspectratio, width=0.75\columnwidth]{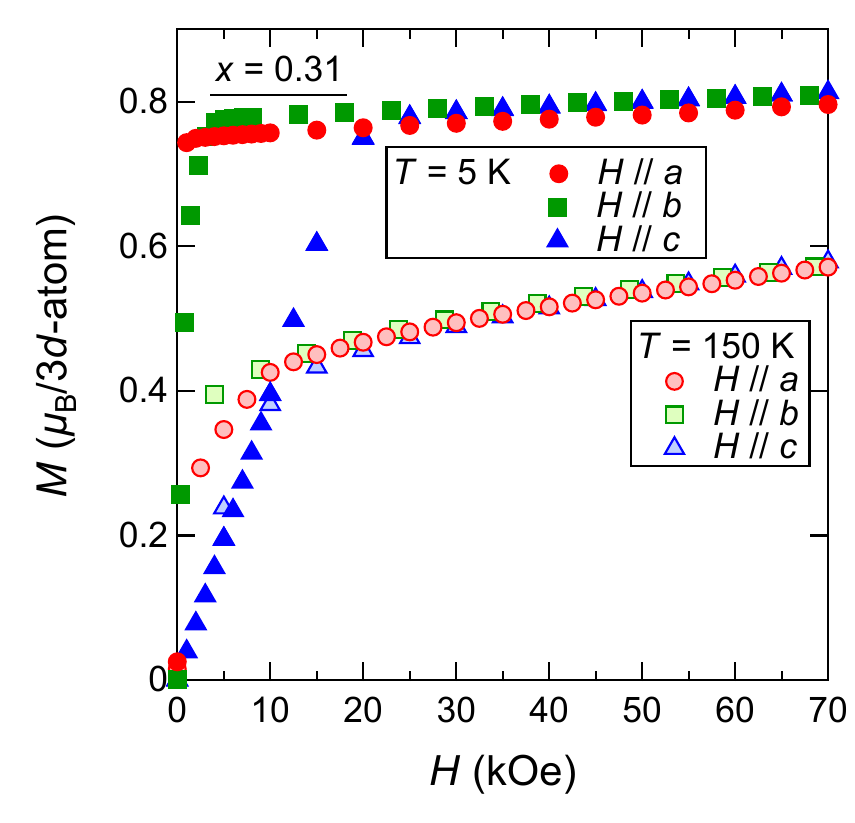}
    \caption{Field dependence of the magnetization for $x=0.31$ at $T=5$ K and 150 K.}
    \label{MHthrsix}
\end{figure}
\indent Figure\ \ref{MHlowx} shows the field dependences of the magnetization for $x=0$--0.46 at $T=5$ K. The magnetic field along the $b$ axis, perpendicular to the plane of the crystal, was corrected to account for the demagnetizing field. The magnetization curves of these samples show typical ferromagnetic behavior. The saturation magnetization of $\Msat =1.25\ \uB$/3$d$-atom for {\FeMAB} decreases with increasing $x$ (see Table\ \ref{transition}). 
The easy magnetization axis of {\FeMAB} is the $a$ axis, and the hard axes are the $b$ and $c$ axes with anisotropy fields of $\Haniso \simeq 10$ kOe and 50 kOe, respectively, which are in agreement with those reported in the previous report \cite{FeSCMAB}. As Mn concentration $x$ increases, the $a$ axis becomes harder and the easy magnetization axis changes to the $b$ axis at $x=0.46$. 
Magnetic anisotropy also changes with temperature. Figure\ \ref{MHthrsix} shows the field dependence of the magnetization for $x=0.31$ at 5 K and 150 K. We can see that the easy magnetization axis is different, the $a$ axis at 5 K and the $b$ axis at 150 K. 
No spin-reorientation transition was suggested from the temperature-dependent magnetization (Fig.\ \ref{MTlowx}). Because it has been reported that the temperature dependence of the magnetocrystalline anisotropy constant of {\xMAB} shows different behavior by the axes \cite{maganisoFe}, the modification of them by Mn substitution is suggested to allow the gradual change of the easy axis with temperature. 
Given the easy magnetization axis along the $b$ axis at $T=5$ K for $x=0.46$, the upper temperature limit of the ferromagnetic phase with the $a$-easy axis is expected to decrease to 0 K with increasing $x$ (see the FMa phase in Fig.\ \ref{Tx}). On the other hand, it is possible that the easy axis rotates in the $a$--$b$ plane with temperature and Mn concentration. The evolution of the magnetic anisotropy at the intermediate $x$ should be determined, for example, by the magnetic torque method. 
As shown in Fig.\ \ref{metalowx}, for $x=0.31$--0.46 the field-dependent magnetization exhibits a rapid increase at a finite magnetic field applied along the $a$ axis, which is parallel to the spin direction in the antiferromagnetic state. The corresponding peak in the $dM/dH$, appears below $\TN$ and shifts to a lower field with decreasing temperature, splitting into doublets at $\Hmone$ and $\Hmtwo$ at low temperatures. With decreasing temperature further, $\Hmone$ and $\Hmtwo$ decrease and reach 0 Oe at $\TC$ and $\TCC$, respectively, and then the anomalies disappear at the lower temperature. 
The peaks of $dM/dH$ are broad and have no hysteresis, which is characteristic of the second-order phase transition. The magnetization is saturated above $\Hmone$, suggesting that at $\Hmtwo <H< \Hmone$ the magnetic structure has an antiferromagnetic component and the spins are rotated towards the ferromagnetic state with increasing field. This should be confirmed by neutron diffraction experiments.
For $x=0.46$, similar transitions were also observed in $dM/dT$ along the $c$ axis (Fig.\ \ref{metalowx}(d)), suggesting that the spins lie along both the axes or in the $a$--$c$ plane. However, it is currently difficult to determine the spin direction of the antiferromagnetic state from the temperature dependence of the magnetization.\\
\begin{figure*}[ht]
\centering
    \begin{minipage}[h]{0.55\linewidth}
    \begin{tabular}{cc}
      \begin{minipage}[h]{0.45\linewidth}
        \centering
        \includegraphics[keepaspectratio, width=\columnwidth]{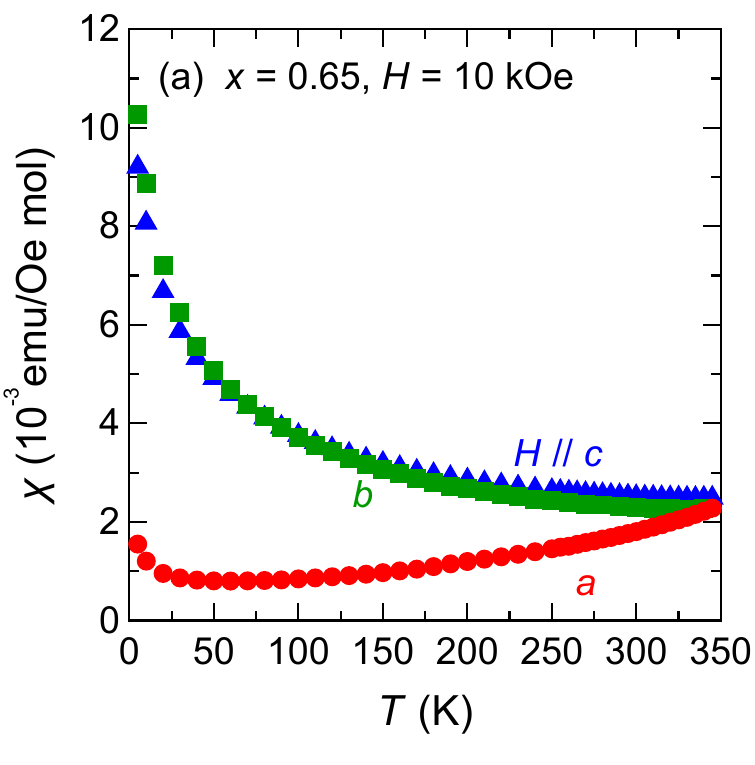}
      \end{minipage}&
      \begin{minipage}[h]{0.45\linewidth}
        \centering
        \includegraphics[keepaspectratio, width=\columnwidth]{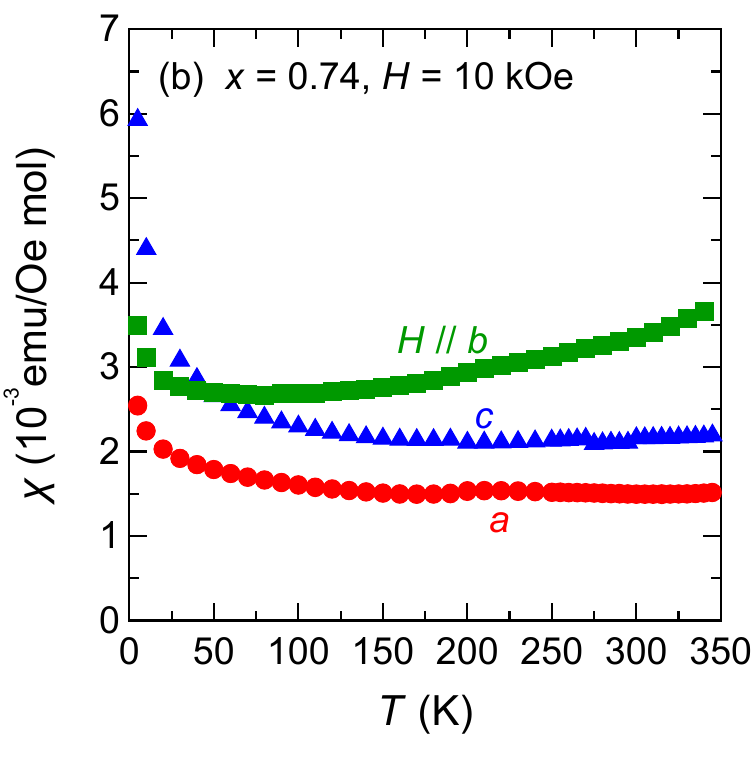}
      \end{minipage}\\
      \begin{minipage}[h]{0.45\linewidth}
        \centering
        \includegraphics[keepaspectratio, width=\columnwidth]{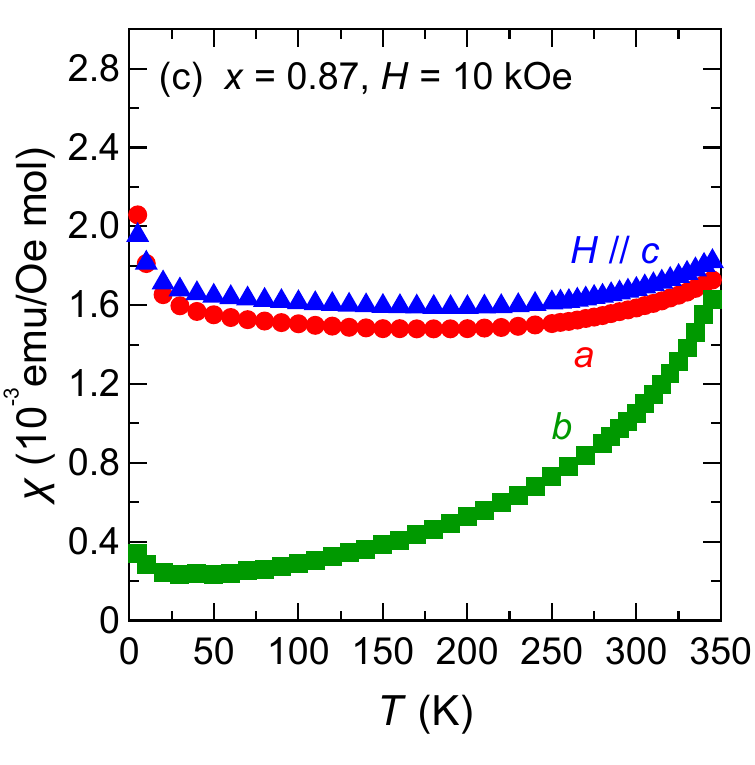}
    \end{minipage}&
    \begin{minipage}[h]{0.45\linewidth}
        \centering
        \includegraphics[keepaspectratio, width=\columnwidth]{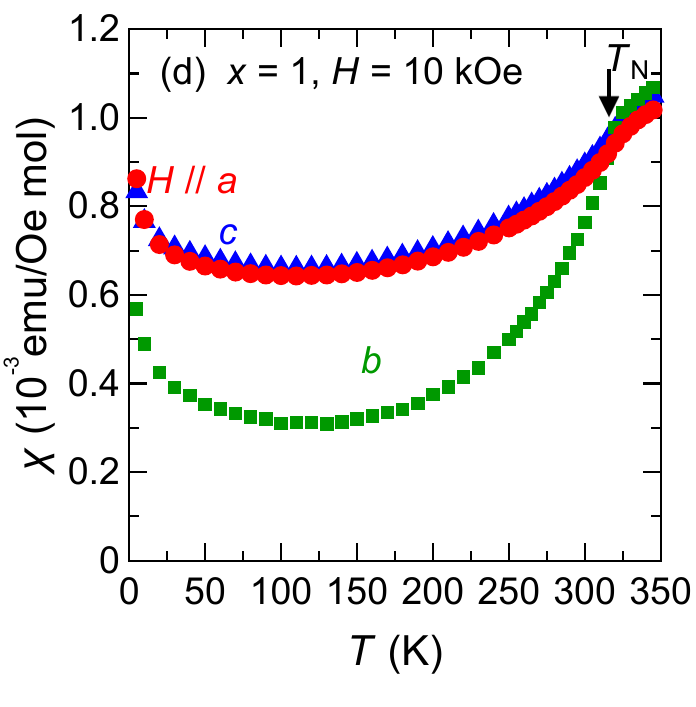}
    \end{minipage}
    \end{tabular}
    \end{minipage}
    \begin{minipage}[h]{0.42\linewidth}
        \centering
        \includegraphics[keepaspectratio, width=\columnwidth]{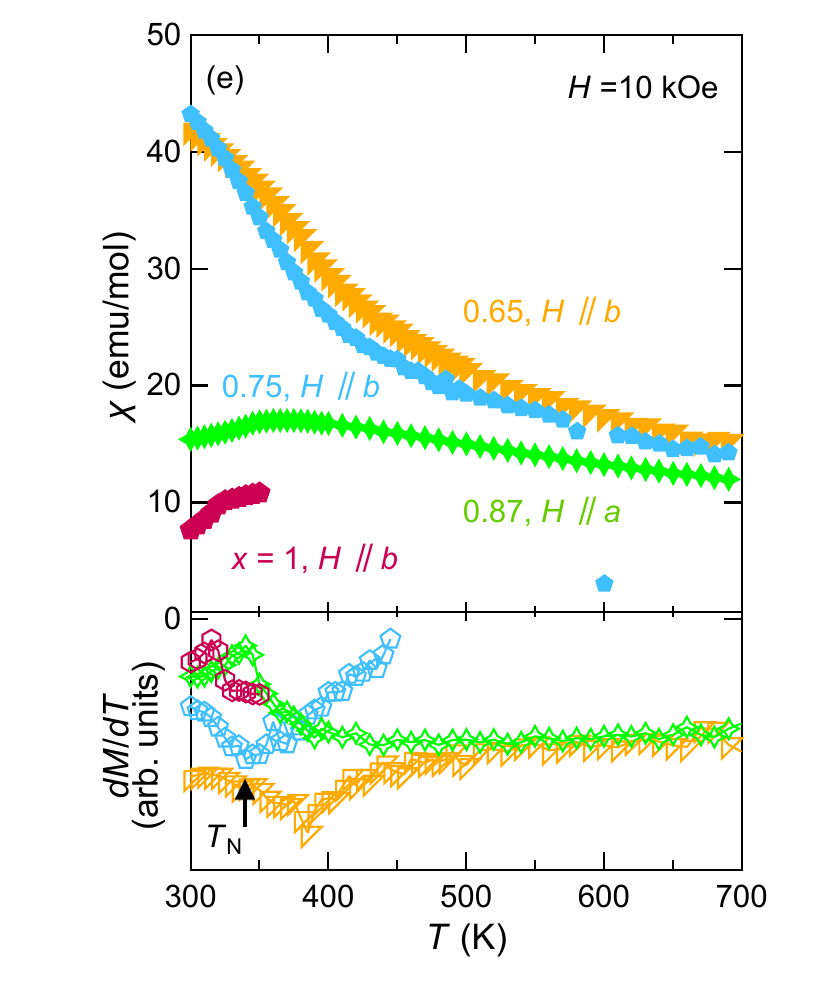}
    \end{minipage}
    \caption{(a)--(d) Temperature dependences of the susceptibility for $x=0.65$--1. (e) The susceptibility for $x=0.65$--0.87 above 300 K measured using the oven option.}
    \label{MThighx}
\end{figure*}
\begin{figure}[ht]
\begin{tabular}{cc}
  \begin{minipage}[h]{0.45\linewidth}
    \centering
    \includegraphics[keepaspectratio, width=\columnwidth]{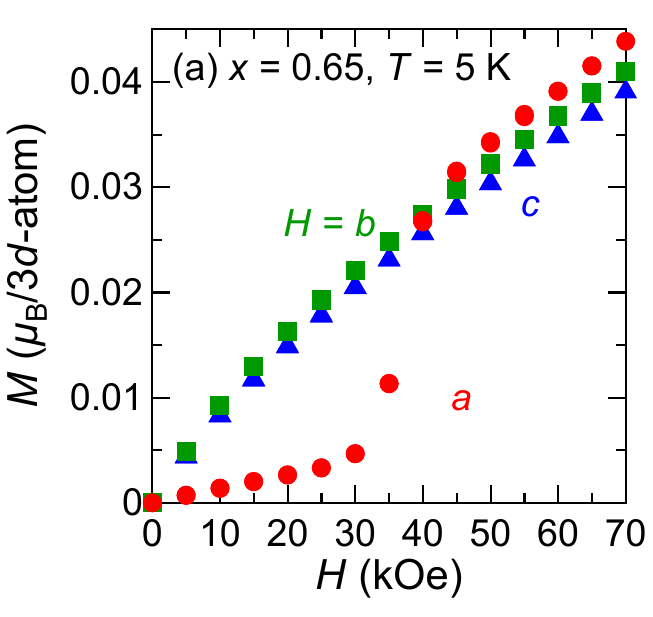}
  \end{minipage}&
    \begin{minipage}[h]{0.45\linewidth}
    \centering
    \includegraphics[keepaspectratio, width=\columnwidth]{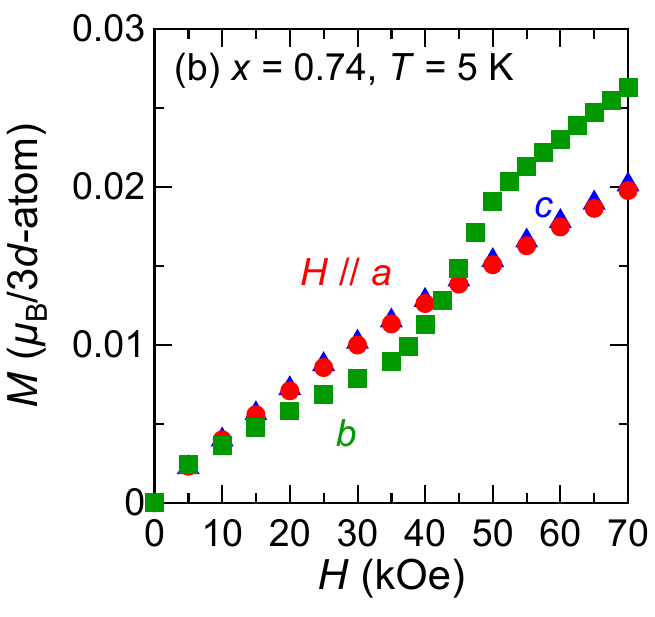}
  \end{minipage}\\
    \begin{minipage}[h]{0.45\linewidth}
    \centering
    \includegraphics[keepaspectratio, width=\columnwidth]{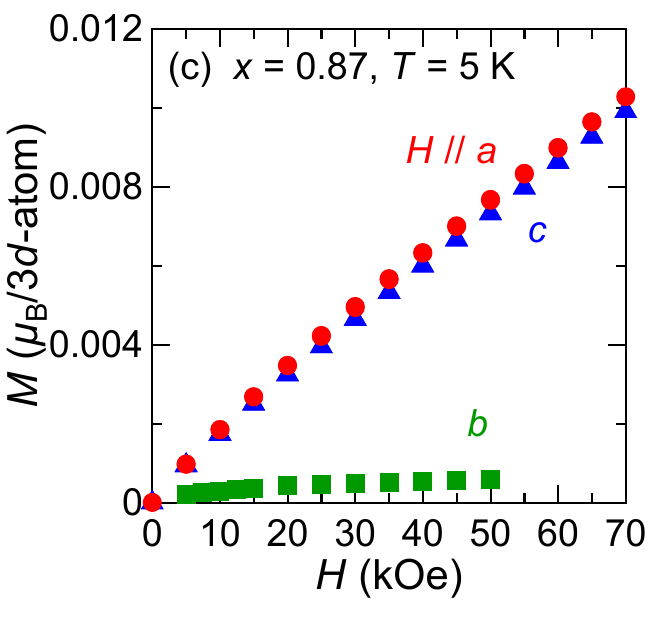}
  \end{minipage}&
    \begin{minipage}[h]{0.45\linewidth}
    \centering
    \includegraphics[keepaspectratio, width=\columnwidth]{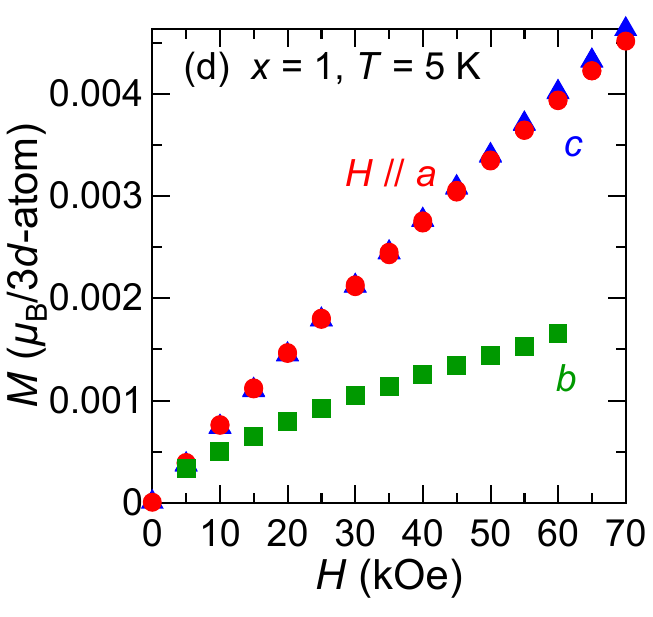}
  \end{minipage}\\
\end{tabular}
  \caption{Magnetization curves for $x=0.65$--1 at 5 K.}
  \label{MHhighx}
\end{figure}
\begin{figure}[ht]
\begin{tabular}{cc}
  \begin{minipage}[h]{0.61\linewidth}
    \centering
    \includegraphics[keepaspectratio, width=\columnwidth]{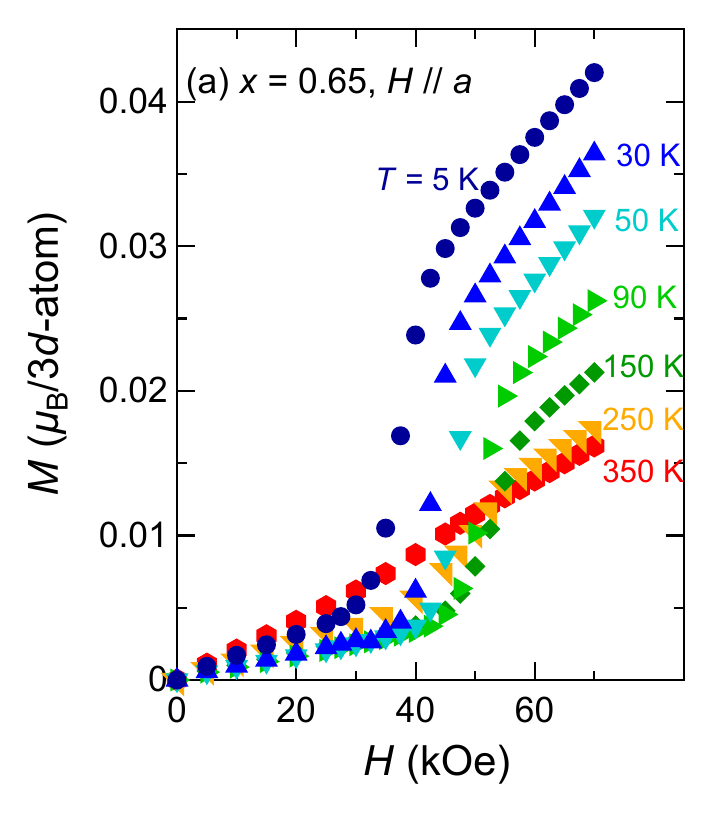}
  \end{minipage}&
    \begin{minipage}[h]{0.28\linewidth}
    \centering
    \includegraphics[keepaspectratio, width=\columnwidth]{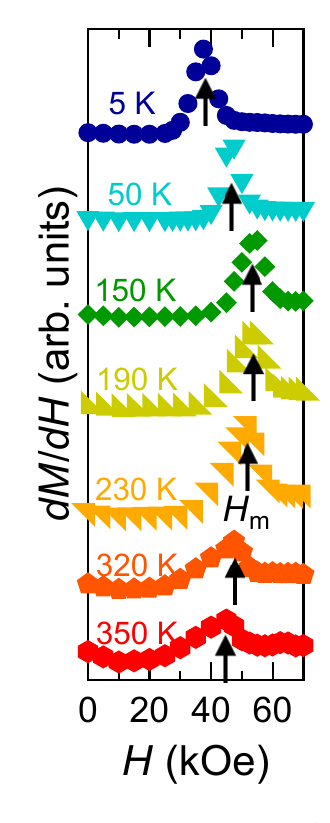}
  \end{minipage}\\
    \begin{minipage}[h]{0.61\linewidth}
    \centering
    \includegraphics[keepaspectratio, width=\columnwidth]{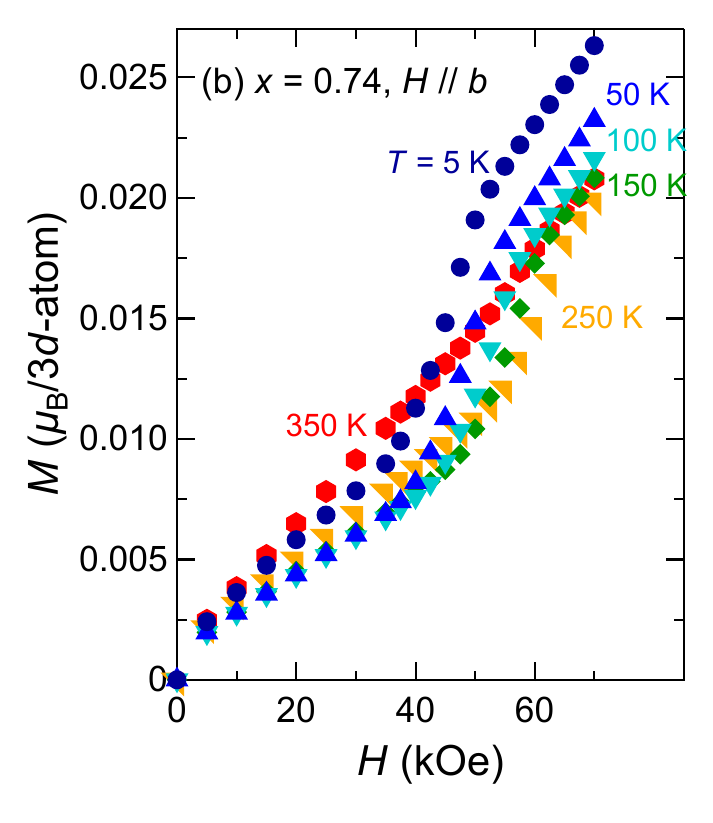}
  \end{minipage}&
    \begin{minipage}[h]{0.28\linewidth}
    \centering
    \includegraphics[keepaspectratio, width=\columnwidth]{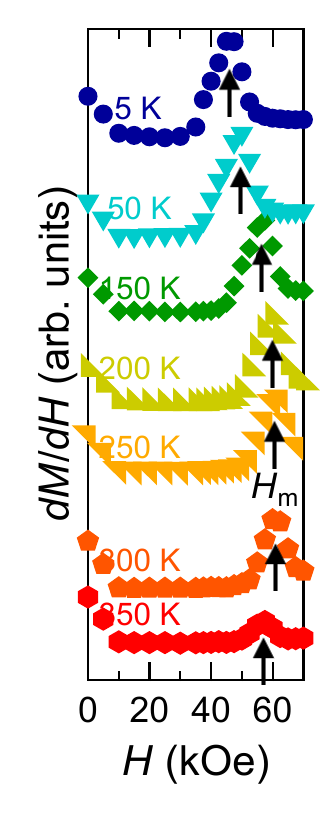}
  \end{minipage}\\
\end{tabular}
  \caption{Magnetization curves and its field derivative at different temperatures (a) parallel to the $a$ axis for $x=0.65$ and (b) to the $b$ axis for $x=0.74$.}
  \label{metahighx}
\end{figure}
\indent Figure\ \ref{MThighx} shows the temperature dependence of the susceptibility measured at 10 kOe for $x = 0.65$--1. These samples show antiferromagnetic behavior. The N\'{e}el temperature of {\MnMAB} is found to be  $\TN = 315$ K, close to the literature data \cite{MnSCMAB}, and increases with decreasing Mn concentration $x$ (Table \ref{transition}), indicating that the antiferromagnetic correlation is enhanced by the Fe substitution. When $x$ decreases to 0.65, the spin direction changes from the $b$ axis to the $a$ axis, which is comparable to the estimated spin direction in the antiferromagnetic phase at low Mn concentrations ($x=0.31$ and 0.36). 
A slight upturn at low temperature was observed for $x=0.87$ and 1, suggesting the presence of a small amount of paramagnetic impurities. On the other hand, for $x=0.65$ and 0.74, Curie-type temperature dependence was observed along particular directions; namely, the $b$ and $c$ axes for $x=0.65$ and the $c$ axis for $x=0.74$. This is ascribed to the enhanced ferromagnetic correlation due to the Fe substitution. \\
\indent Figure\ \ref{MHhighx} shows the field dependence of the magnetization for $x=0.65$--1 at $T=5$ K. The magnetization increases linearly along all the axes except for the $a$ axis for $x=0.65$ and the $b$ axis for $x=0.74$, where a metamagnetic-like transition was observed. 
Figure\ \ref{metahighx} shows the field-dependent magnetization and its derivative $dM/dH$ measured at different temperatures for $x=0.65$ and 0.74. The metamagnetic-like transition was observed at the field parallel to the spin direction in the whole measured temperature range of $T=5$--350 K, and probably up to $\TN$. Above the transition field $\Hm$, the magnetization is not saturated but increases linearly, which is characteristic of the spin-flop process. 
In general, a metamagnetic transition occurs with field hysteresis and a first-order discontinuous increase in magnetization. On the other hand, at the transition, the magnetization increased rapidly but continuously, and $dM/dH$ shows a broad peak even at $T = 5$ K. Therefore, this metamagnetic-like transition is considered to be of the gradual spin-flop type, which occurs in the case of weak magnetocrystalline anisotropy \cite{SF, SFnpj, SFErMnSn, SFCuMnS, SFGdCo}.
In a uniaxial antiferromagnet, the metamagnetic field can be simply written as $\Hm =\sqrt{2\HA \HE - \HA^2}$, where $\HA$ and $\HE$ are the magnitudes of the anisotropy field and the antiferromagnetic exchange field, respectively.  
Given the near-room-temperature antiferromagnetic ordering, the exchange field $\HE$ is likely to be large and thus the anisotropy field $\HA$ should be small in order to induce the transition at the relatively low fields (Fig.\ \ref{metahighx}). 
This conjecture agrees with the condition of the gradual-spin-flop transition, and one of the proposed scenarios for the small $\HA$ is a competition of uniaxial magnetic anisotropy of the $a$ and $b$ axes. 
Moreover, $\Hm$ takes a maximum at an intermediate temperature below $\TN$. This may be ascribed to a competition between the magnetic anisotropy and the antiferromagnetic correlation.  
\section{Magnetic phase diagrams}
\begin{figure}[ht]
    \centering
    \includegraphics[keepaspectratio, width=0.9\columnwidth]{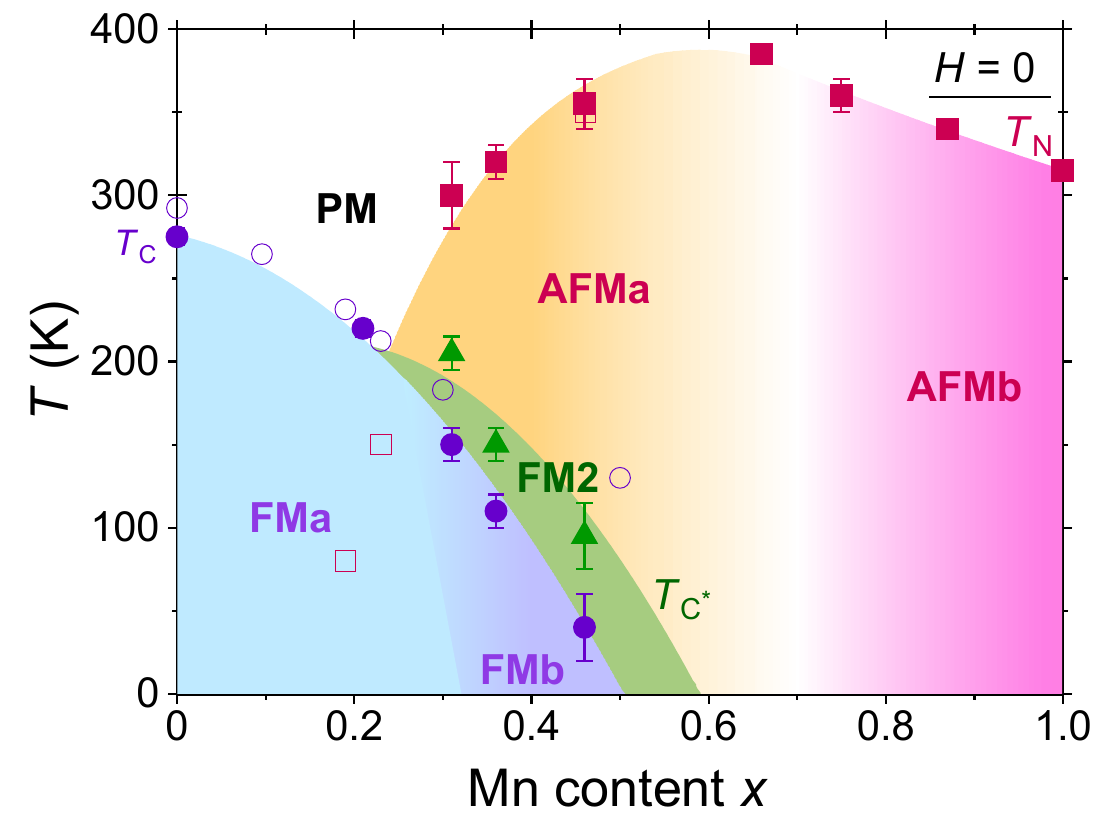}
    \caption{$T$--$x$ magnetic phase diagram at zero field for {\xMAB}. Open makers represent transition temperatures estimated from neutron powder diffraction experiments \cite{MnFePRM}. }
    \label{Tx}
\end{figure}
Based on the results of magnetization measurements, we tried to construct magnetic phase diagrams of {\xMAB}. 
Figure\ \ref{Tx} shows the $T$--$x$ magnetic phase diagram. It consists of three distinct magnetic phases; ferromagnetic (FM) phases and antiferromagnetic (AFM) phases, and an intermediate (FM2) phase. 
In the FM phase the easy magnetization axis changes with temperature and Mn concentration; FM with the $a$-easy-magnetization axis (FMa) is in the low $x$ region while FM with $b$-easy-magnetization axis (FMb) is close to the FM2 phase. In the AFM phase, the spin direction changes from the $b$ axis (AFMb) to the $a$ axis (AFMa) below $x \sim 0.7$. 
The propagation vector is expected to be ${\bm q} = (0,0,1/2)$ as given by the the previous neutron powder diffraction measurements \cite{MnFePRM, MABMnmagpowd}. Although Potashnikov \textit{et al}. \cite{MnFePRM} reported an antiferromagnetic transition below $\TC$ at $x=0.19$ and 0.23 (see open markers in Fig.\ \ref{Tx}), our magnetization data show no corresponding anomaly for $x<0.21$. Note that the powder sample used for the neutron diffraction measurements contained impurity phases, which may lead to extrinsic magnetic diffraction. Otherwise, the observed transition temperatures are approximately close to the literature data \cite{MnFePRM}.\\
\indent The N\'{e}el temperature $\TN$ has a maximum at around $x=0.6$ and drops rapidly below $x=0.6$ while the Curie temperature $\TC$ decreases monotonically with increasing $x$. The Curie temperature was found to split into two transitions at $\TC$ and $\TCC$ at around $x=0.2$, leading to the intermediate FM2 phase. The FM2 phase is surrounded by the FMb and AFMa phases. 
In this region, a rapid increase of magnetization was observed under the finite field along the $a$ axis (Fig.\ \ref{metalowx}). Therefore, the FM2 phase is considered to have both a ferromagnetic component along the $b$ axis and a small antiferromagnetic component along the $a$ axis. The magnetic state is comparable to the canted antiferromagnetic structure with ${\bm q} = (0,0,1/2)$ predicted by neutron powder diffraction experiments \cite{MnFePRM}. Neutron diffraction measurements on single crystals are future work.
These complex variations of the spin direction at the intermediate Mn concentration seem to be difficult to be explained by the rigid-band model. Lu {\etal} suggested that the 3$d$ orbitals, which contribute to the high density of states $D(E)$ and the nearly flat bands near the Fermi level $E_{\rm F}$, are different between {\FeMAB} ($d_{xy}$) and {\MnMAB} ($e_g$), and give rise to the different spin directions \cite{MnDOS}. Thus, it is possible that the intermediate nature of the $3d$-band structure near $E_{\rm F}$ makes the uniaxial magnetic anisotropy competitive in {\xMAB}.   
In Fig.\ \ref{Tx}, there seems to be a quadruple critical point around ($x$, $T$) = (0.2, 200 K). However, systematic studies using samples with more finely controlled compositions are needed to elucidate its nature.\\
\begin{figure*}[ht]
\begin{tabular}{ccc}
  \begin{minipage}[h]{0.32\linewidth}
    \centering
    \includegraphics[keepaspectratio, width=\columnwidth]{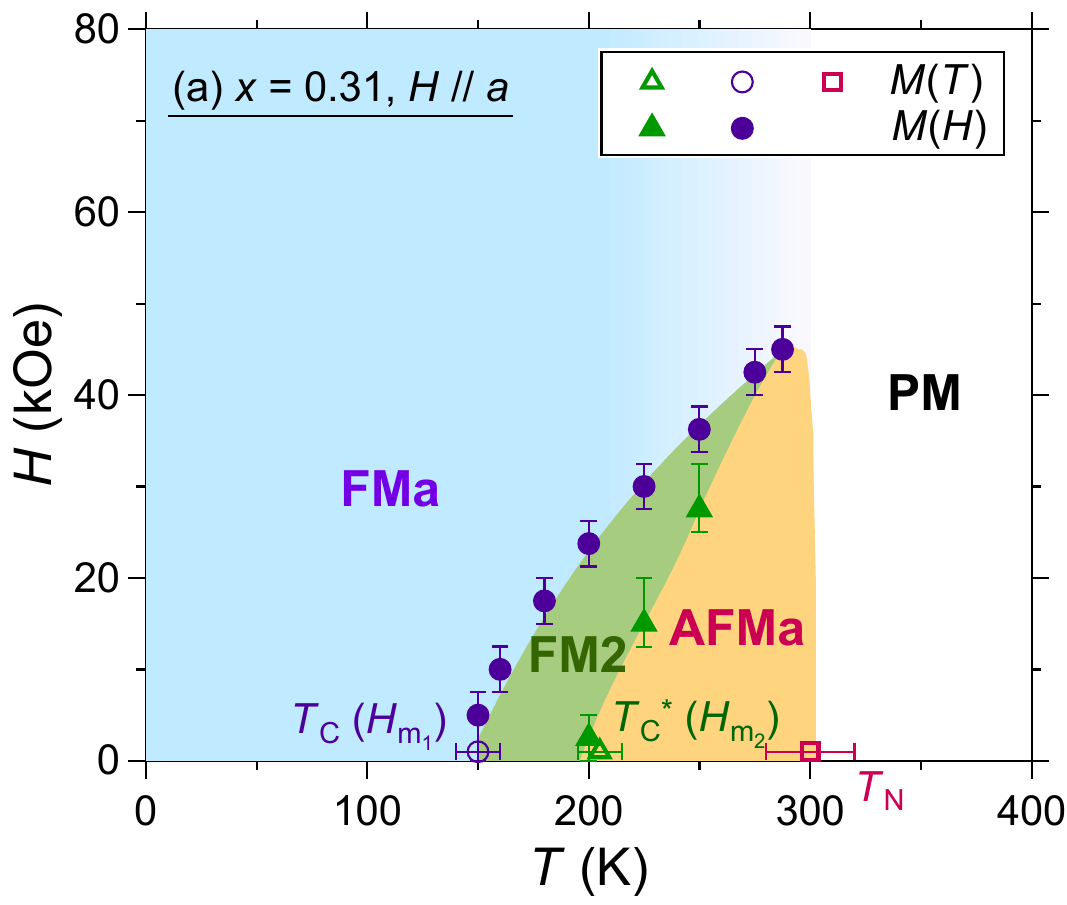}
  \end{minipage}&
    \begin{minipage}[h]{0.32\linewidth}
    \centering
    \includegraphics[keepaspectratio, width=\columnwidth]{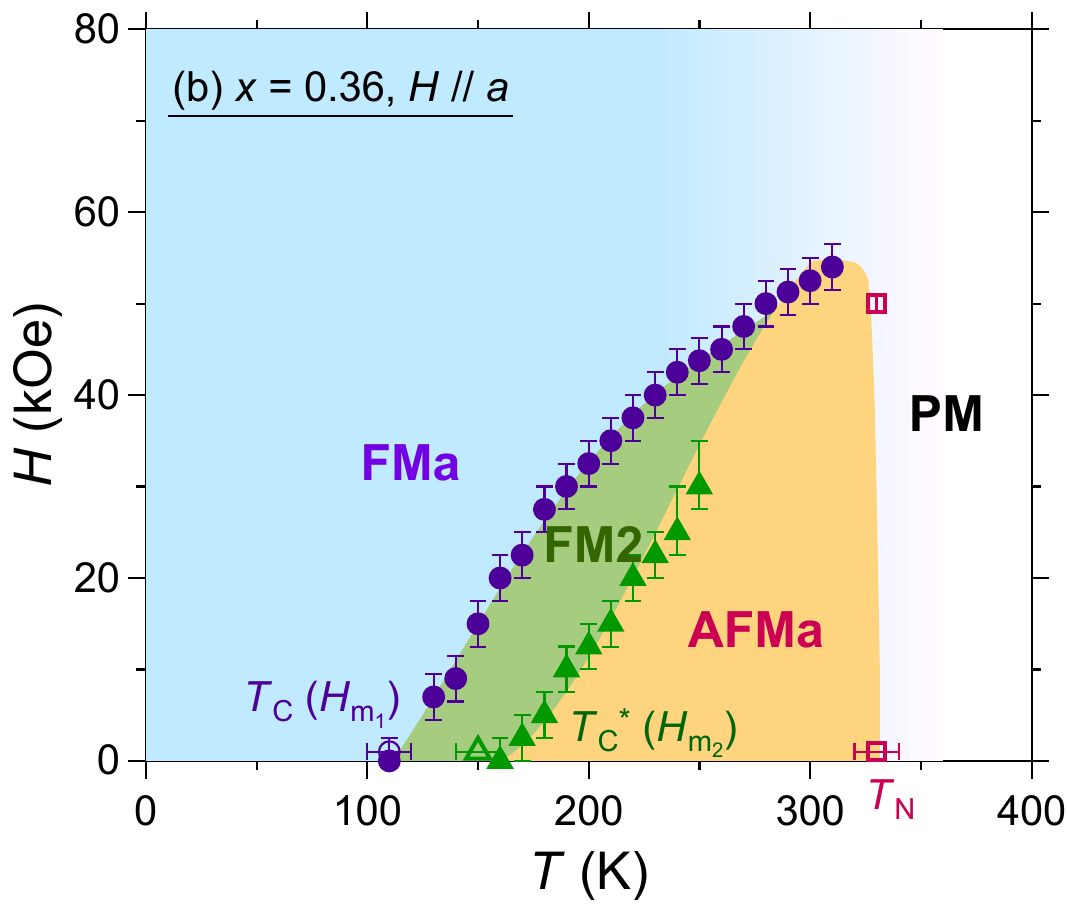}
  \end{minipage}&
    \begin{minipage}[h]{0.32\linewidth}
    \centering
    \includegraphics[keepaspectratio, width=\columnwidth]{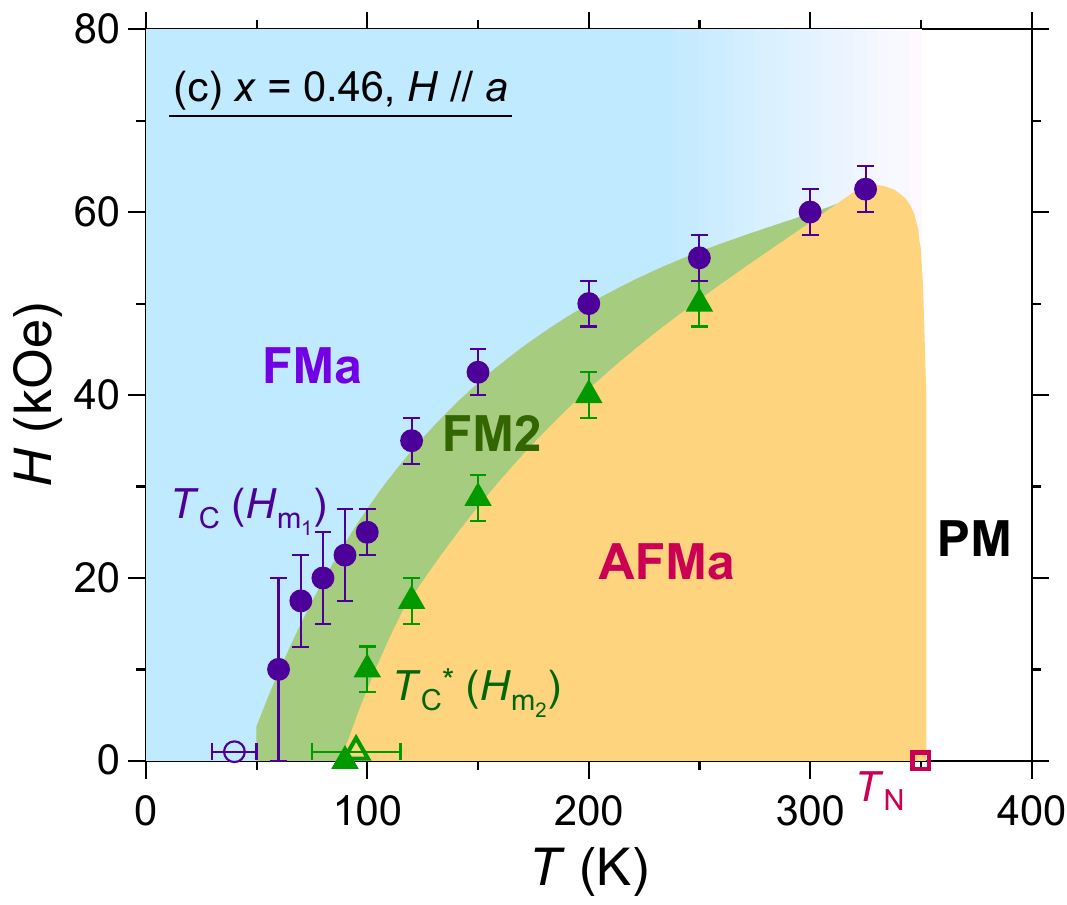}
  \end{minipage}\\
\end{tabular}
  \caption{$H$--$T$ magnetic phase diagrams for $x=0.31$--0.46 under the field applied to the $a$ axis. Closed and open marks represent the data obtained from the field-dependent and temperature-dependent magnetization, respectively.}
  \label{HTlow}
\end{figure*}
\begin{figure}[ht]
  \begin{tabular}{c}
    \begin{minipage}[h]{0.9\linewidth}
      \centering
      \includegraphics[keepaspectratio, width=\columnwidth]{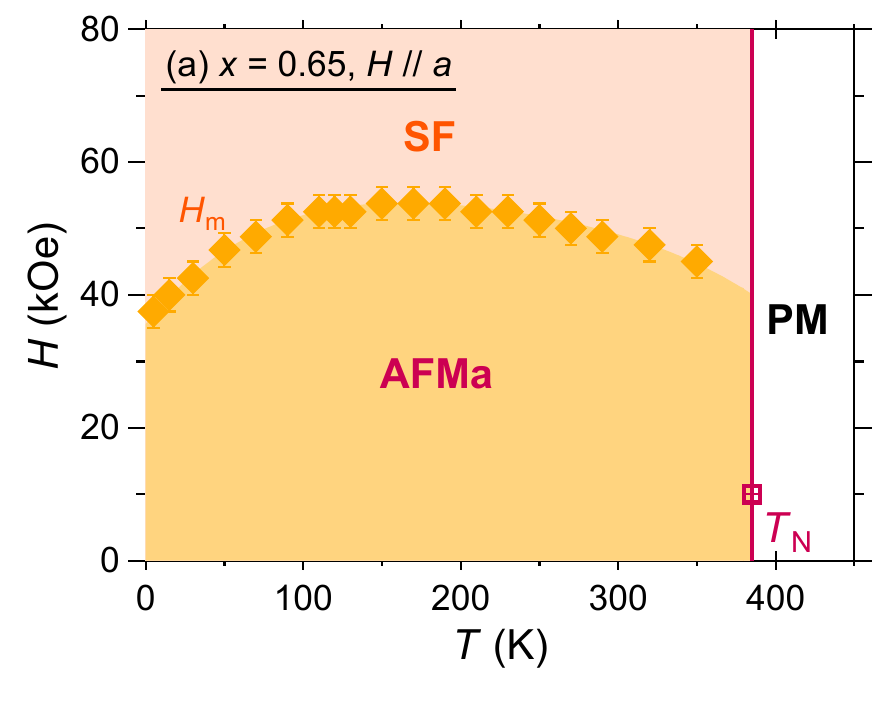}
    \end{minipage}\\
      \begin{minipage}[h]{0.9\linewidth}
      \centering
      \includegraphics[keepaspectratio, width=\columnwidth]{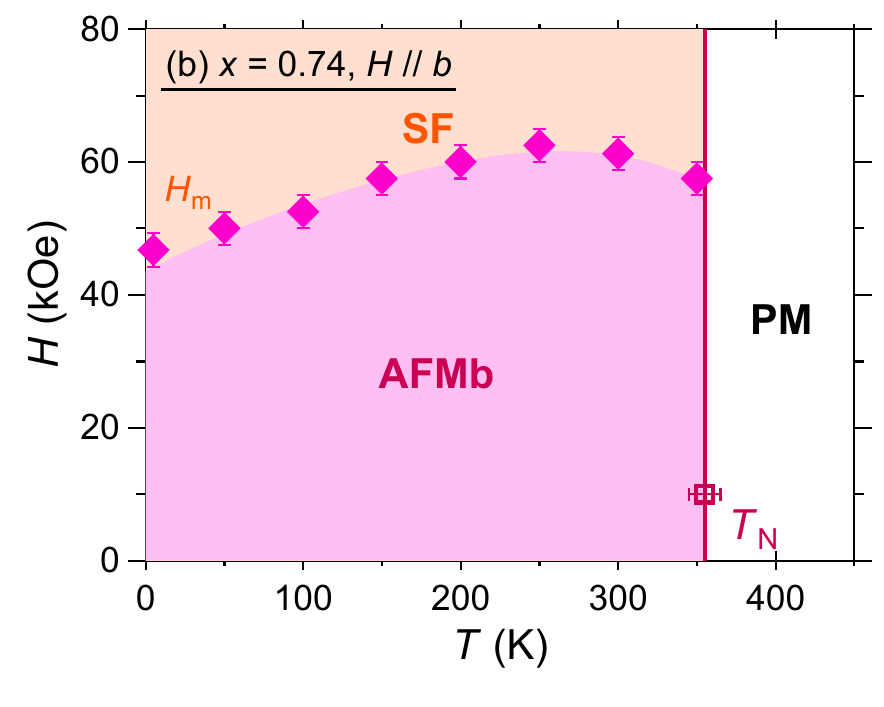}
    \end{minipage}\\
  \end{tabular}
    \caption{$H$--$T$ magnetic phase diagrams for (a) $x=0.65$ and (b) 0.74 under the field along the $a$ and $b$ axes, respectively.}
    \label{HThigh}
  \end{figure}
\begin{figure}[ht]
    \centering
    \includegraphics[keepaspectratio, width=1\columnwidth]{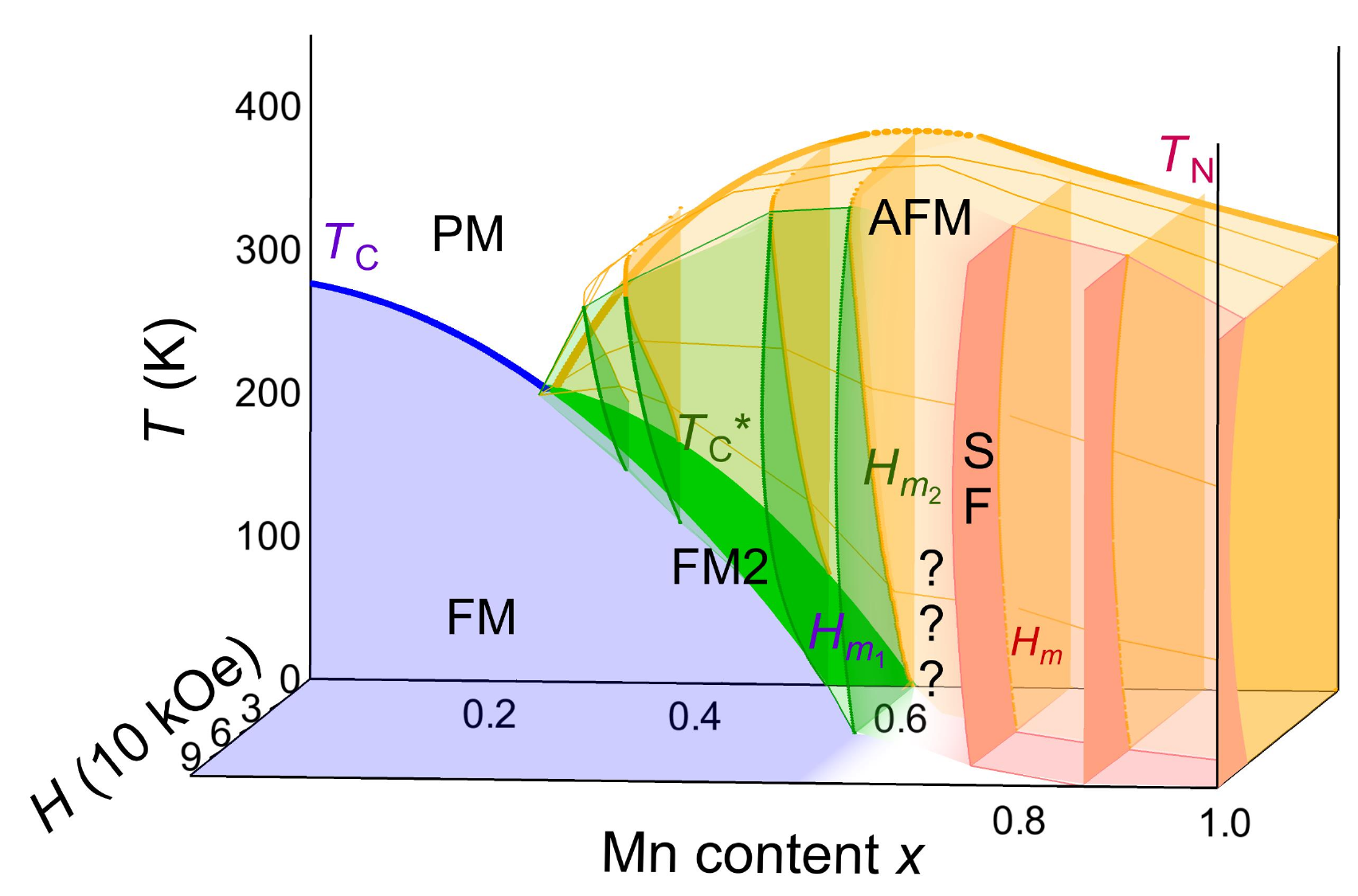}
    \caption{Three-dimensional magnetic phase diagram with applied magnetic field, temperature and Mn contnent.}
    \label{xTH}
\end{figure}
\indent From the temperature-evolution of the anomalies observed in magnetic isotherm measured under the field along the $a$ axis for $x=0.31$, 0.36, and 0.46, we constructed a $H$--$T$ magnetic phase diagram (Figs.\ \ref{HTlow}(a)--(c)). Note that in the case of $x=0.46$, similar anomalies were also observed along the $c$ axis and we obtained almost the same phase diagram as along the $a$ axis. 
The AFMa phase is located in the high-temperature region and expands as Mn concentration increases. It undergoes two-step transitions to a field-induced ferromagnetic phase with increasing field. Given that the transition field $\Hmone$ and $\Hmtwo$ decrease to zero at around $\TC$ and $\TCC$, respectively, the intermediate phase in Fig.\ \ref{HTlow} is considered to be identical to the FM2 phase in the $T$--$x$ diagram (Fig.\ \ref{Tx}). \\
\indent Figure\ \ref{HThigh} shows the $H$--$T$ magnetic phase diagram under the field applied to the $a$ and $b$ axes for $x=0.65$ and 0.74, respectively. The spin-flop (SF) phases are located above the dome-shaped AFM phase below $\TN$. The transition between the SF and the AFM phases is gradual and continuous. This is ascribed to the weak uniaxial magnetic anisotropy competing between along the $a$ and $b$ axes. This may lead to the spin rotation in the $a$--$b$ plane. 
A field-induced ferromagnetic state was not reached in the field range of 0--70 kOe. A higher field is necessary to know whether the SF state undergoes another antiferromagnetic phase transition or saturates to a ferromagnetic state.\\
\indent Finally, we obtained the three-dimensional $H$--$T$--$x$ magnetic phase diagram (Fig.\ \ref{xTH}). For simplicity, the spin direction is not included. In the $x=0.2$--0.6 region, the AFM phase is robust to magnetic fields in the high temperature region because it is far from the FM phase. With increasing $x$, the critical field $\Hmtwo$ near $\TN$ becomes higher, and the transition temperature $\TCC$, where $\Hmtwo$ reaches 0 Oe, decreases due to the suppressed ferromagnetic and enhanced antiferromagnetic correlations. Between AFM and FM phases, there is an intermediate phase FM2, and its region extends similarly with increasing $x$.
Above $x = 0.6$, where the ferromagnetic correlation disappears, only AFM phase is present below $\TN$ at low fields, and it rapidly becomes robust to magnetic fields even for the low temperature region. 
Under fields higher than $\Hm$, a spin flop phase SF appears. Given that it arises with a dominant antiferromagnetic component as a result of a competition of uniaxial magnetic anisotropy, its character is different from the FM2 phase with a dominant ferromagnetic component, which arises from a competition between ferromagnetic and antiferromagnetic correlations. Although the details are unclear at present, it is expected that the SF phase will lie below the FM2 phase (i.e. $\Hm < \Hmtwo$) or appear as a drastic change in the character of the FM2 phase (i.e. $\Hmtwo$ is identical to $\Hm$) at $x \simeq 0.6$.

\section{Conclusion}
Single crystals of nanolaminated borides {\xMAB} were synthesized in the entire Fe--Mn composition range using the self-flux method, and structural and magnetization measurements were performed. 
The Curie temperature of $\TC \simeq 275\ $K and the spontaneous moment of $\Msat \simeq 1.3\ \uB/$Fe in {\FeMAB} decrease monotonically with increasing Mn concentration, whereas the N\'{e}el temperature has a maximum of $\TN \simeq 385\ $K at $x$ = 0.65.
The spin direction in the ferromagnetic and antiferromagnetic states at 5 K changed from the $a$ axis to the $b$ axis as the Mn concentration increased to $x \simeq 0.4$ and 0.7, respectively. 
In the range of $x$ = 0.31--0.46, we observed an antiferromagnetic ordering above $\TCC$ and two-step ferromagnetic transitions at $\TC$ and $\TCC$ along the $a$ axis.
In this intermediate phase FM2, both the ferromagnetic correlations of the $b$-axis spin component and the antiferromagnetic correlations of the $a$-axis spin component coexist.
At $x$ = 0.65 and 0.74, a gradual spin-flop transition was observed below $\TN$ in the field along the $a$ and $b$ axes, respectively, due to the weak uniaxial magnetic anisotropy caused by the competition between the $a$ and $b$ axes.
The magnetic properties of {\xMAB} are therefore found to be sensitive to the composition, temperature and field. Our findings could further expand the variety of potential applications as an environment-resistant itinerant magnet.\\
\begin{acknowledgments}
The authors thank Y. Uno for their help with composition analysis. This work was supported by SPRING, Support for Pioneering Research Initiated by the Next Generation of Kyoto University.
\end{acknowledgments}

\bibliography{ref}

\end{document}